\documentclass[lettersize,journal]{IEEEtran}
\usepackage{amsmath,amsfonts, amssymb,bm,amsthm}
\usepackage{algorithm}
\usepackage{algorithmic}
\usepackage{array}
\usepackage[caption=false,font=normalsize,labelfont=sf,textfont=sf]{subfig}
\usepackage{textcomp}
\usepackage{stfloats}
\usepackage{url}
\usepackage{verbatim}
\usepackage{cite}
\usepackage{multirow}
\usepackage{graphicx}

\def\thline{\noalign{\hrule height 1pt}}

\def\thline{\noalign{\hrule height 1pt}}

\newtheorem{theorem}{Theorem}
\newtheorem{Rem}{Remark}
\def\thline{\noalign{\hrule height 1pt}}

\def\thline{\noalign{\hrule height 1pt}}

\newcommand{\argmin}{\mathop{\rm argmin}\limits}
\def\BibTeX{{\rm B\kern-.05em{\sc i\kern-.025em b}\kern-.08em
    T\kern-.1667em\lower.7ex\hbox{E}\kern-.125emX}}
\usepackage{balance}

\begin{document}
\title{A Design of Denser-Graph-Frequency Graph Fourier Frames for Graph Signal Analysis}
\author{Kaito Nitani and Seisuke Kyochi~\IEEEmembership{Member,~IEEE,}
\thanks{K. Nitani and S. Kyochi are with Kogakuin University.}
\thanks{This work was supported by JSPS Grants-in-Aid (24K07481) and ROIS NII Open Collaborative Research 2024 (24S0108).}
\thanks{Manuscript received XXX XX, XXXX.}}
\markboth{Journal of \LaTeX\ Class Files,~Vol.~18, No.~9, September~2020}%
{How to Use the IEEEtran \LaTeX \ Templates}

\maketitle

\begin{abstract}
This paper introduces a design method for denser-graph-frequency graph Fourier frames (DGFFs) to enhance graph signal processing and analysis. The graph Fourier transform (GFT) enables us to analyze graph signals in the graph spectral domain and facilitates various graph signal processing tasks, such as filtering, sampling and reconstruction, denoising, and so on. However, the conventional GFT faces two significant limitations. First, unlike the discrete Fourier transform and its variants (such as discrete cosine transforms), the graph frequencies of the derived graph Fourier basis (GFB) from a given graph tend to be unevenly distributed or localized, which leads to biased spectral analysis. Second, the GFB used in GFT does not provide an efficient sparse representation of graph signals compared to overcomplete systems like frames. To overcome these challenges, we propose adding oscillating vectors with intermediate graph frequencies between the original vectors in the GFB for both undirected and directed graphs, constructing GFFs with denser-graph frequencies. The resulting DGFFs are expected to enable more accurate graph signal analysis. Furthermore, we propose a graph filtering method based on the DGFFs. In experiments, we apply the DGFFs to practical applications such as graph signal recovery, demonstrating superior performance compared to existing GFBs.
\end{abstract}


\section{Introduction}
\IEEEPARstart{G}{raph} signal processing (GSP) has emerged as an essential framework for analyzing signals defined on irregular structures represented by graphs. Traditional signal processing handles signals in regular domains, such as time or space, while GSP focuses on signals indexed by graph nodes \cite{Ortega_2022, GSP1, GSP2, GSP3, GSP4, GSP5, GSP6, GSP7, GSP8}. Depending on the application, graph structures may be undirected or directed, where nodes represent objects such as sensors, pixels, or individuals, and edges encode their relationships \cite{DirectedGSP}. GSP extends classical concepts like filtering \cite{filter1, filter2, filter3, TFfilter, OSGFB, dabush2024verifying, isufi2024graph, kalofolias2016learn}, sampling \cite{sampling1, sampling2, sampling3}, and domain transforms (e.g., frequency domain), offering new ways to extract and analyze graph-embedded information.

A key GSP technique is the graph Fourier transform (GFT), which generalizes the Fourier transform to graph signals. In undirected graphs, the GFT derives from the eigenvalue decomposition of the graph Laplacian, enabling frequency analysis. However, for directed graphs, the asymmetry of the directed Laplacian complicates eigenvalue decomposition, making traditional computation of the graph Fourier basis (GFB) difficult. To address this, several alternative methods have been proposed \cite{MagGFT, AuGFT, SVDGFT, PDGFT, PDfilter, GDV, DGFT, HRWGFT, GenBC, StableApp, DiLap, DAGs}, summarized in Sec. \ref{sec:GFT}.

\begin{figure}[t]
 \centering
         \begin{minipage}[b]{0.48\linewidth}
 			\centering
 				\scalebox{0.17}{\includegraphics[keepaspectratio=true]{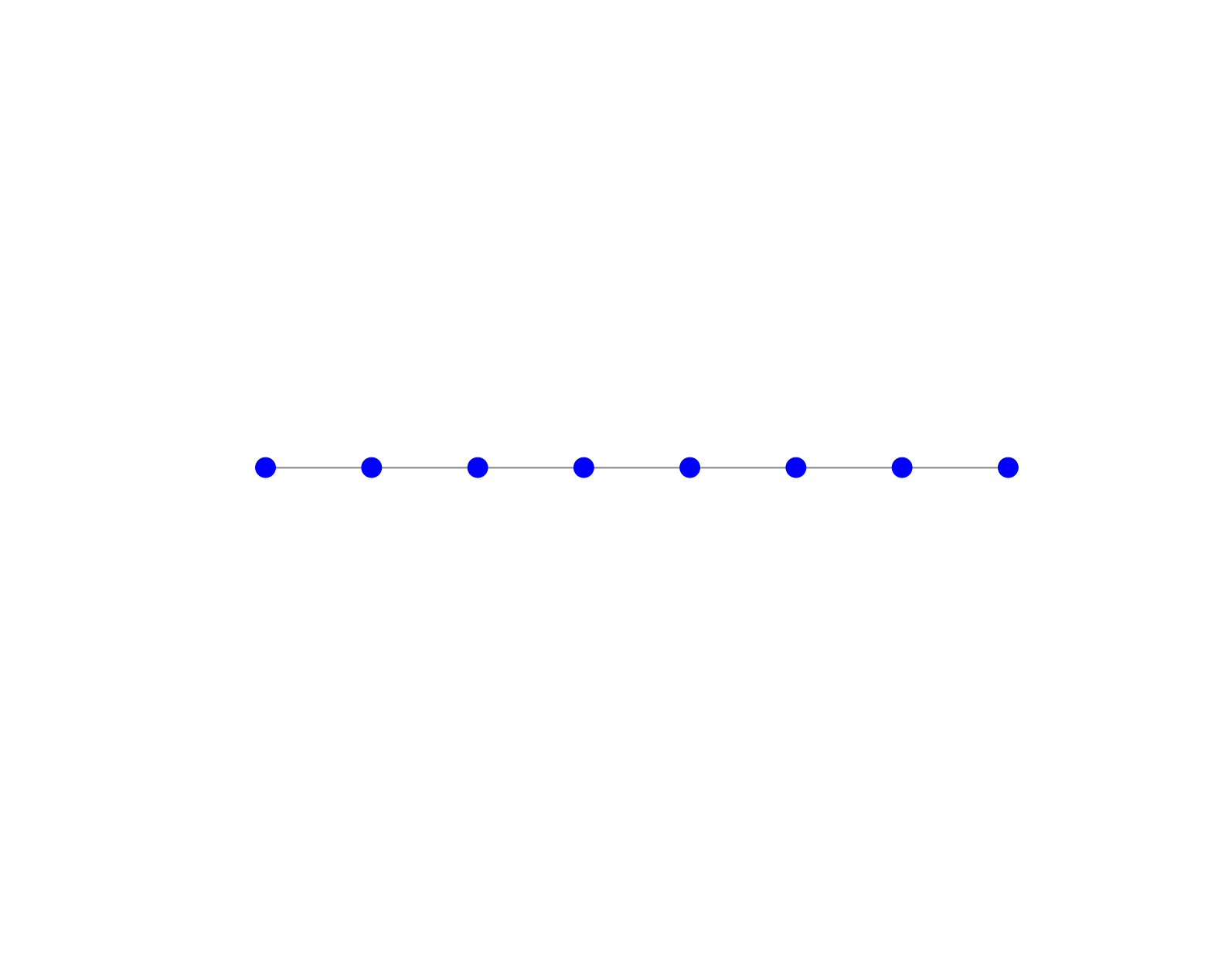}}
 				\centerline{(a) Path graph }
 	\end{minipage}
  			\begin{minipage}[b]{0.48\linewidth}
 			\centering
 				\scalebox{0.17}{\includegraphics[keepaspectratio=true]{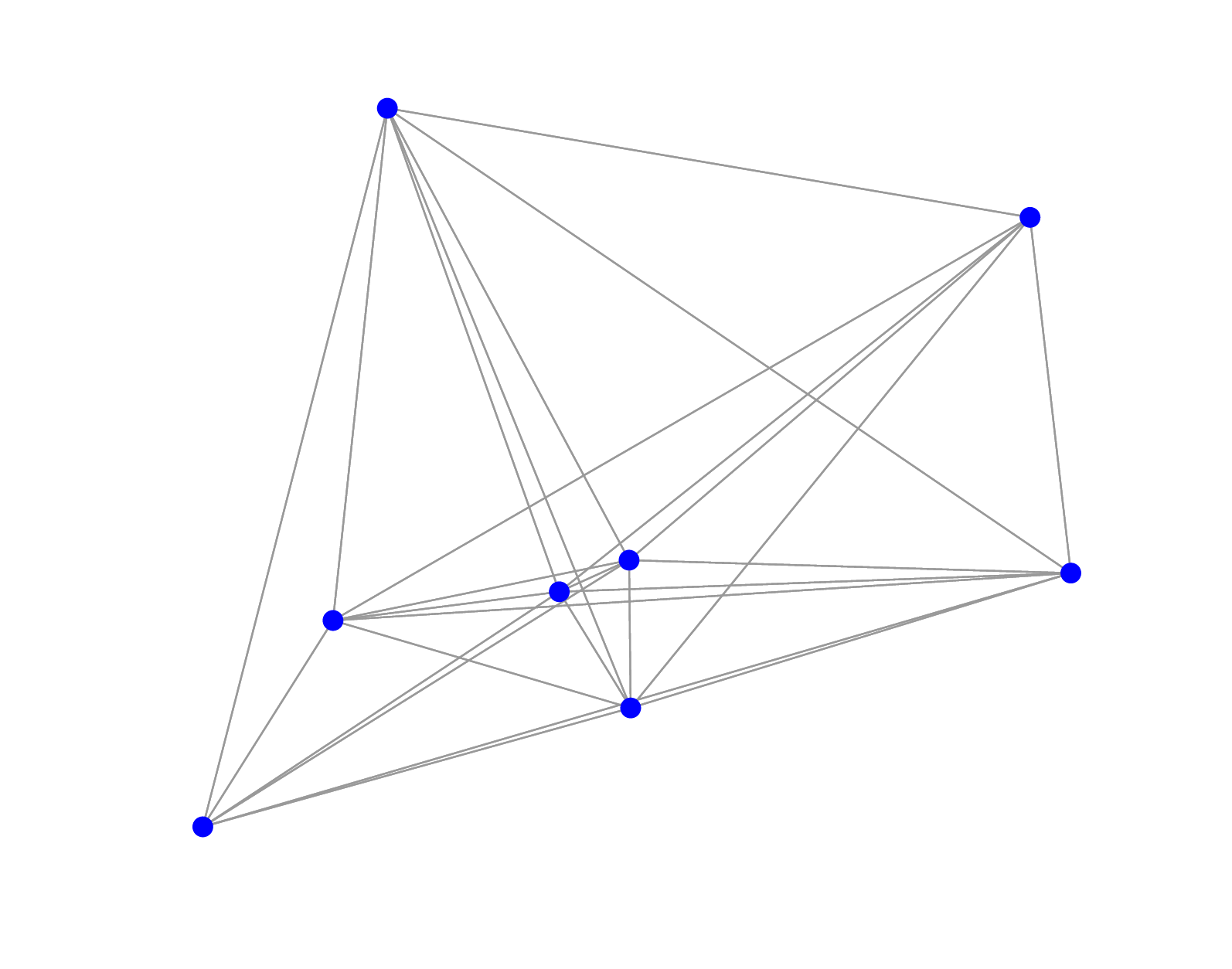}}
 				\centerline{(b) Sensor graph }
 	\end{minipage}
	
 			\begin{minipage}[b]{0.48\linewidth}
 			\centering
 				\scalebox{0.17}{\includegraphics[keepaspectratio=true]{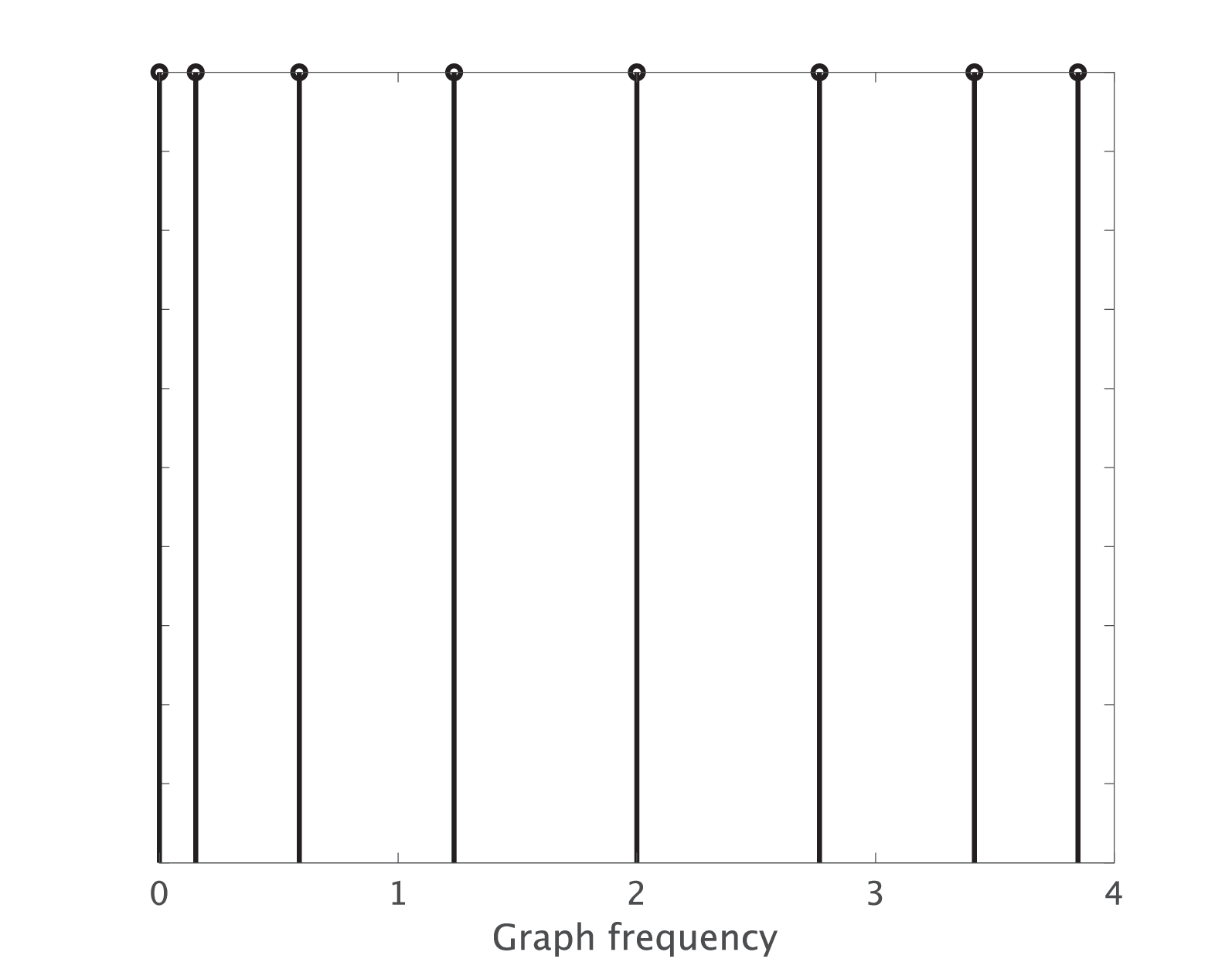}}
 				\centerline{(c) GF of path graph}
 	\end{minipage}
  			\begin{minipage}[b]{0.48\linewidth}
 			\centering
 				\scalebox{0.17}{\includegraphics[keepaspectratio=true]{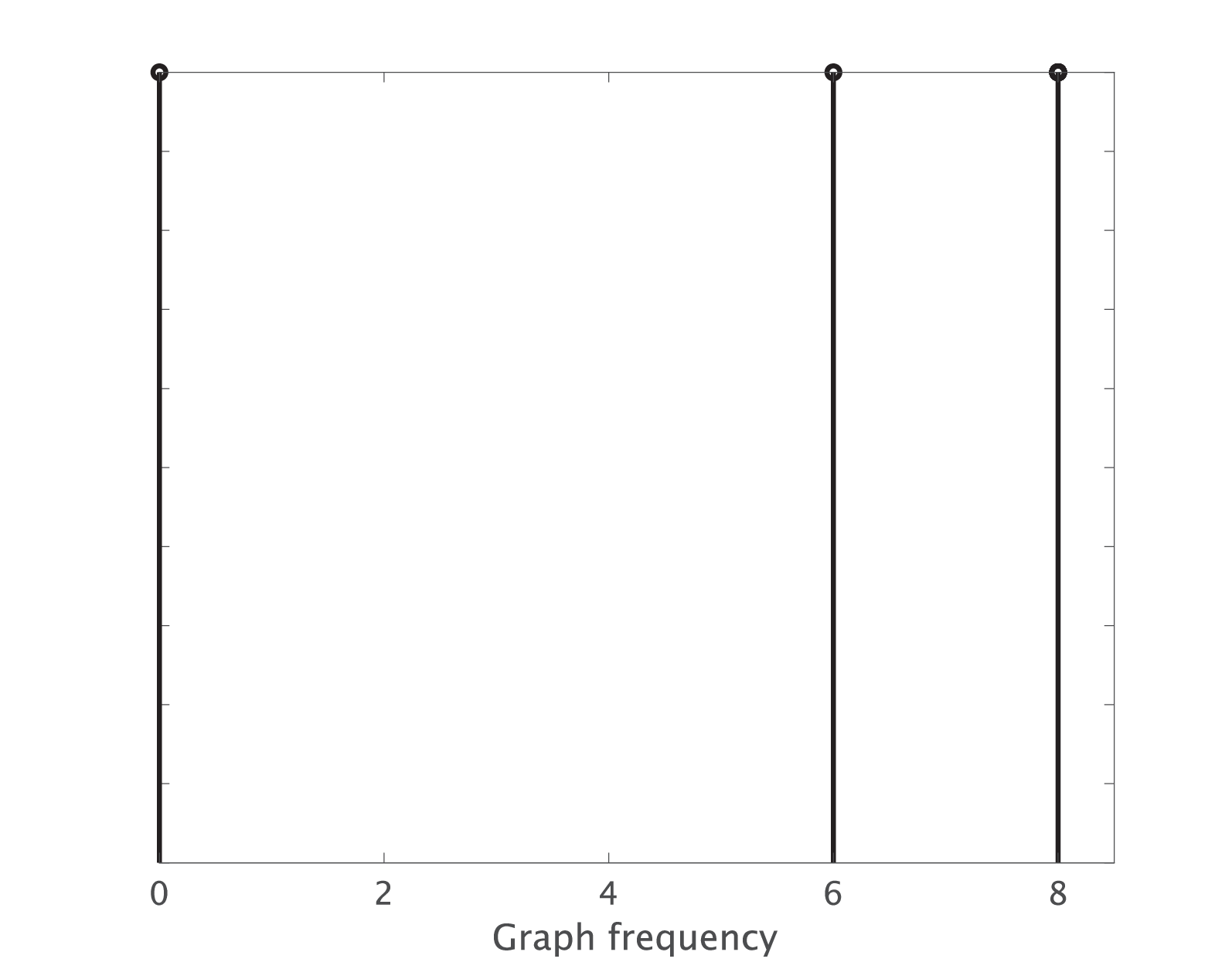}}
 				\centerline{(d) GF of sensor graph}
 	\end{minipage}
 	\caption{
 Graphs and graph frequencies (GFs) of GFB (Num. of nodes: $N=8$)}\label{fig:recim1}
 \vspace{-0.3cm}
 \end{figure}

Despite defining the GFT, these methods face two major limitations. First, the graph frequency distribution is often uneven or localized. For example, Fig. \ref{fig:recim1} compares graph frequencies of undirected path and sensor graphs, revealing a severe bias in the sensor graph. This uneven distribution leads to a biased spectral representation, overrepresenting some frequencies while underrepresenting others. A GFT addressing this issue via an optimization problem under orthogonality constraints was discussed in \cite{DGFT}. Second, the GFT-based GFB  is not the most suitable for efficient sparse representation (SR) of graph signals. In contrast, frames (or dictionaries) are generally more effective for SR \cite{Mallat2008, Rubinstein2010, Elad2010}.

In this study, to address both non-uniform graph frequency distribution and sparse representation (SR) limitations in conventional GFTs, we propose \textit{denser-graph-frequency graph Fourier frames} (DGFFs)\footnote{A finite or countably infinite set of vectors $\{\mathbf{f}_n\}_{n \in \Lambda} \subset \mathcal{H}$ is a \textit{frame} for a Hilbert space $\mathcal{H}$ with inner product $\langle \cdot , \cdot \rangle$ if there exist constants $\alpha_{\mathrm{min}} > 0$ and $\alpha_{\mathrm{max}} < \infty$ such that  
\begin{align*}
    \alpha_{\mathrm{min}} \| \mathbf{x} \|^2 \leq \sum_{n \in \Lambda } |\langle \mathbf{x} , \mathbf{f}_n \rangle |^2 \leq \alpha_{\mathrm{max}}\|\mathbf{x}\|^2, \quad \forall \mathbf{x} \in \mathcal{H}.
\end{align*}
Here, we restrict $\mathcal{H}$ to $\mathbb{R}^N$ and consider overcomplete sets $\{\mathbf{f}_m\}_{m=1}^M$ ($N < M$) with $\mathrm{rank}([\mathbf{f}_m]_{m=1}^M) = N$, ensuring they form frames.}. DGFFs extend bases into frames by adding vectors with intermediate graph frequencies absent in the initial GFB, mitigating biased spectral analysis and enhancing SR capability.

\subsection{Contributions}
This paper contributes as follows:
\begin{enumerate}
    \item{DGFFs for undirected graphs:} We propose two methods. Some GFBs are analytically represented by continuous functions, e.g., the discrete Fourier basis (DFB) for ring graphs and the discrete cosine basis (DCB) for path graphs. For such cases, we introduce \textit{Analytic DGFF} (ADGFF), generating oscillating wave vectors with intermediate graph frequencies. Second, we propose \textit{Linear-interpolation-based DGFFs} (LiDGFFs), where undirected GFBs, formed by eigenvectors of the graph Laplacian, allow additional oscillating vectors via linear combinations of adjacent vectors, constructing LiDGFFs.
    \item{Low-redundant DGFFs:} Graphs may have regions with sparse or dense graph frequency distributions, where dense regions do not require extra oscillating vectors. To reduce redundancy, we introduce \textit{low-redundant LiDGFFs} (lrLiDGFFs), improving computational efficiency in practical applications.
    \item {DGFFs for directed graphs:} Some GFBs are derived from optimization problems rather than eigenvalue decomposition, e.g., the directed GFB in \cite{DGFT} via directed variation (DV) minimization. For such cases, we propose \textit{Optimization-based DGFFs} (OptDGFFs), constructing frames from intermediate graph frequencies through optimization based on DV.
    \item{Graph spectral filtering:} Filtering is a key tool in signal analysis, extensively studied in classical signal processing and GSP. In GSP, {graph spectral filtering (GS filtering)} operates in the graph spectral domain. We propose \textit{denser-graph-spectral filtering} (DGS filtering) based on DGFFs, enabling finer frequency component filtering.
\end{enumerate}

\begin{table*}[t]
\caption{Proposed DGFFs}
\label{tab:compCSHT}
\begin{center}
\begin{tabular}[c]{c|c|c|c|c|c|c}
\thline
Category & GFBs & ADGFFs & LiDGFFs & lrLiDGFFs & OptDGFFs & DGS filtering \\\hline
\multirow{2}{*}{Analytic GFB} & DFB (GFB on ring graph) &  \checkmark &  \checkmark  &   ------ &  \checkmark  & \checkmark   \\\cline{2-7}
                                  & DCB (GFB on path graph) &  \checkmark &  \checkmark &   ------ &  \checkmark  & \checkmark  \\\hline
\multirow{2}{*}{EVD-based GFB} & (Undirected) GFB&   ------ & \checkmark &   \checkmark &  \checkmark  & \checkmark \\\cline{2-7}
                                    & (Directed) Magnetic GFB \cite{MagGFT} &   ------ & \checkmark &   \checkmark  &  \checkmark  & \checkmark  
                                       \\\hline
\multirow{1}{*}{Opt-based GFB} & (Directed) SfGFB \cite{DGFT} & ------ & ------  &   ------& \checkmark & \checkmark  \\ \thline                                
\end{tabular}
\end{center}
\end{table*}

Table \ref{tab:compCSHT} summarizes the proposed methods in this work, and shows which existing GFTs each method is applicable to. The preliminary work of this study is introduced in \cite{Nitani2024}, where we designed only the LiDGFF for undirected graphs. In contrast, this paper further extends the design method for the LiDGFF to directed graph signals and proposes the ADGFF, the lrLiDGFFs, the OptDGFF, and DGS filtering for both undirected and directed graphs. 

\subsection{Paper Organization}
The remainder of this paper is organized as follows. In Sec. \ref{sec:GFT}, we provide an overview of GFT tools for both undirected and directed graphs, along with their corresponding GFBs and variation measures. Sec. \ref{sec:DGFF} introduces the design method of DGFFs for each type of GFB. Additionally, in Sec. \ref{sec:DGFFfiltering}, we extend graph spectral filtering based on the GFB to DGS filtering based on the DGFFs. The proposed DGFFs are evaluated in Sec. \ref{sec:experimental} through experiments on the DGS filtering performance and graph signal recovery through SR. Finally, this paper is concluded in Sec. \ref{sec:conclusion}.

\subsection{Notations} \label{subsec:notation}
Bold-faced lower-case letters and upper-case letters are vectors and matrices, respectively. Other mathematical notations are summarized in Table \ref{tab:Notations}. 

\begin{table}[t]
\caption{Basic notations}
\vspace{-0.2cm}
\label{tab:Notations}
\begin{center}
\scalebox{0.75}{ 
\begin{tabular}{c|c}
\thline
Notation & Terminology \\ \thline
$\mathbb{R}$, and $\mathbb{R}_+$& real, nonnegative real numbers \\ \hline
$\mathbb{C}$, $\mathrm{j}$ & Complex numbers and complex unit $\sqrt{-1}$ \\ \hline
${A}^N$ and ${A}^{M\times N}$ ($A \subset \mathbb{C}$)& \begin{tabular}{c}
 $N$- and $M \times N$-dimensional vectors/matrices \\  with elements in $A$
\end{tabular} \\ \hline
$\mathbf{I}$, $\mathbf{J}$, $\mathbf{O}$& Identity, reversal identity, and zero matrix \\ \hline
$\mathbf{X}^{\top}$, $\mathbf{X}^{\mathrm{H}}$ & Transpose of $\mathbf{X}$, and conjugate transpose of $\mathbf{X}$ \\ \hline
$x_n$ and $[\mathbf{x}]_n$& $n$-th element of a vector $\mathbf{x}$ \\ \hline
$X_{m,n}$ and $[\mathbf{X}]_{m,n}$& $(m,n)$-th element of a matrix $\mathbf{X}$ \\ \hline
$[\mathbf{x}_{n}]_{n=1}^M \in A^{N \times M}$ $(\mathbf{x}_n \in A^N)$ & Matrix consisting of $N$ vectors $\mathbf{X} = \begin{bmatrix}
    \mathbf{x}_1 & \cdots & \mathbf{x}_N
\end{bmatrix}$ \\ \hline
\begin{tabular}{c}$\mathrm{Diag}(\mathbf{x}) = \mathrm{Diag}(x_1, \ldots , x_{N})$, \\
 $\mathrm{Diag}( \mathbf{X}^{(\mathrm{1})}, \ldots , \mathbf{X}^{({N})})$
 \end{tabular}
 & Diagonal/block-diagonal matrices \\ \hline
 $\mathrm{diag}(\mathbf{X}) \in A^M$ 
 & Diagonal elements of $\mathbf{X} \in A^{M\times N}$ \\ \hline
  $\mathrm{Sym}(\mathbf{X}) := \frac{1}{2}(\mathbf{X} + \mathbf{X}^\top)$ 
 & Symmetrization of $\mathbf{X} \in A^{N\times N}$ \\ \hline
$\|\mathbf{x}\|_p$ ($p \in [1 , \infty )$)  & $\ell_p$-norm, $\|\mathbf{x}\|_p = \left(\sum_{n=1}^{N} |x_n|^p\right)^{\frac{1}{p}}$ \\ \hline
$\mathcal{B}_2(\mathbf{y},\epsilon)$  & \begin{tabular}{c}
    $\mathbf{y}$-centered $\ell_2$-norm ball with radius $\epsilon \in \mathbb{R}_+$ \\
$\mathcal{B}_2(\mathbf{y},\epsilon) = \{ \mathbf{x} \in \mathbb{R}^N\ | \ \|\mathbf{x} - \mathbf{y}\|_2 \leq \epsilon \}$
\end{tabular} \\ \thline
\end{tabular}
}
\end{center}
\vspace{-0.4cm}
\end{table}

\section{Graph Fourier Basis/Transform and Related Topics}\label{sec:GFT}
First, we review the GFB and the GFT, which are fundamental tools for graph signal analysis, along with the graph variation measures associated with the GFB. This section also covers the GFTs for ring graphs, path graphs, and other undirected/directed graphs. For more detailed information on the GFT, please refer to \cite{Ortega_2022, GSP1, GSP2, GSP3, GSP4, GSP5, GSP6, DirectedGSP}.

\subsection{Undirected/Directed Graphs and Undirected GFTs}
A graph $\mathcal{G} = ( \mathcal{V} , \mathcal{E} )$ is defined by a set of vertices $\mathcal{V} = \{v_n\}_{n=1}^{N}$ and a set of edges $\mathcal{E}$, i.e., the collection of connected index pairs. An adjacency matrix $\mathbf{W} \in \mathbb{R}_+^{N\times N}$ encodes the edge weights $\{w_{n_1,n_2}\}_{n_1,n_2 = 1}^{N}$, where $w_{n_1,n_2}=0$ indicates that the $n_1$-th and $n_2$-th nodes are disconnected. We refer to graphs with a symmetric adjacency matrix, i.e., $\mathbf{W} = \mathbf{W}^\top$, as \textit{undirected graphs}, and otherwise as \textit{directed graphs}. 

Based on the adjacency matrix, the GFB for undirected graphs can be derived as follows. Let $\mathbf{D} = \mathrm{Diag}(d_{1},\ldots,d_{N})$ be the degree matrix, where the $n_1$-th diagonal element is defined as $d_{n_1} = \sum_{n_2 = 1}^{N} w_{n_1,n_2}$. Let $\mathbf{L}$ be the graph Laplacian, expressed as $\mathbf{L} = \mathbf{D} - \mathbf{W}$, which satisfies the symmetry property $\mathbf{L} = \mathbf{L}^\top$. The symmetric normalized graph Laplacian is given by $\widetilde{\mathbf{L}}  = \mathbf{D}^{-\frac{1}{2}} \mathbf{L} \mathbf{D}^{-\frac{1}{2}}$. The GFB is derived via eigenvalue decomposition:
\begin{align}
\mathbf{L} = \mathbf{U}\boldsymbol{\Lambda}\mathbf{U}^\top \Longleftrightarrow \mathbf{u}_{k}^\top \mathbf{L} \mathbf{u}_k = \lambda_k.
\end{align}
Similarly, the GFB can also be derived from the eigenvalue decomposition of the normalized graph Laplacian, with its eigenvalues satisfying $0 \leq \widetilde{\lambda}_k \leq 2$. The eigenvalues $\{{\lambda_k}\}_{k=1}^{N}$ are referred to as graph frequencies, and the eigenvectors $\{{\mathbf{u}_k}\}_{k=1}^{N}$ serve as the GFB. For a given graph signal $\mathbf{s} \in \mathbb{R}^N$, the GFT (for undirected graphs) is defined as $\hat{\mathbf{s}} = \mathbf{U}^\top \mathbf{s}$, where the GFT coefficients $ \hat{\mathbf{s}}$ represent the amplitude of the graph signal in each frequency component. 

The eigenvectors of the graph Laplacian for undirected graphs can be interpreted as the solution to an optimization problem that seeks to minimize the graph total variation (GTV) $\mathcal{GTV} : \mathbb{R}^N \rightarrow \mathbb{R}_{++}$ as follows:
\begin{align}\label{TV}
\mathcal{GTV}(\mathbf{x}) :=&\ \mathbf{x}^\top \mathbf{L}\mathbf{x} = \frac{1}{2}\sum\limits_{ n_1,n_2 = 1}^{N} w_{n_1,n_2} (x_{n_1} - x_{n_2})^2,\nonumber\\
\mathbf{u}_k =&\  \argmin_{\substack{\mathbf{x} \perp \{\mathbf{u}_1, \ldots, \mathbf{u}_{k-1}\},\  \|\mathbf{x}\|_2 = 1}} \mathcal{GTV}(\mathbf{x}).
\end{align}
The eigenvector corresponding to the smallest eigenvalue $\lambda_1 = 0$ is the constant vector $\mathbf{u}_1 = \frac{1}{\sqrt{N}}\mathbf{1}$, i.e., $\mathcal{GTV}(\mathbf{u}_1) = 0$.

For directed graphs with a diagonalizable graph Laplacian $\mathbf{L}$, the GFB and GFT (\textit{directed GFB/GFT}) can also be derived using its eigenvectors $\mathbf{L} = \mathbf{U}\boldsymbol{\Lambda}\mathbf{U}^{-1}$. 

\subsection{Analytic GFTs}
For certain undirected/directed graphs, their GFBs can be analytically represented as continuous wave functions. For instance, the DFB $[\mathbf{u}_{k+1}^{(\mathrm{r})}]_{k=0}^{N-1} \in \mathbb{C}^{N \times N}$ and the (type-II) DCB $[\mathbf{u}_{k+1}^{(\mathrm{p})}]_{k=0}^{N-1} \in \mathbb{R}^{N \times N}$ are defined as:
\begin{align}
[\mathbf{u}_{k+1}^{(\mathrm{r})}]_n :=&\ {\frac{1}{\sqrt{N}}}\exp\left[\mathrm{j}\left(\frac{2\pi}{N}k\right)n\right], \\
[\mathbf{u}_{k+1}^{(\mathrm{p})}]_n :=&\ c_k \sqrt{\frac{2}{N}} \cos\left[\left(\frac{\pi}{N}k\right)\left(n+\frac{1}{2}\right)\right],\\
c_k =&\ \tfrac{1}{\sqrt{2}}\ (k=0),\ c_k = 1 \ (k\neq 0),
\end{align}
which are the GFBs of \textit{(directed) ring graphs} and \textit{path graphs}, respectively. Their eigenvalues are given by:
\begin{align}
\mathbf{L}^{(\mathrm{r})} \mathbf{u}_k^{(\mathrm{r})} =&\  \lambda_k^{(\mathrm{r})} \mathbf{u}_k^{(\mathrm{r})},\ \ \lambda_k^{(\mathrm{r})} = 1 - \exp\left[\mathrm{j}\frac{2\pi}{N}k\right],\\
\mathbf{L}^{(\mathrm{p})} \mathbf{u}_k^{(\mathrm{p})} =&\ \lambda_k^{(\mathrm{p})} \mathbf{u}_k^{(\mathrm{p})},\ \ \lambda_k^{(\mathrm{p})} = 2 - \cos\left[\frac{\pi}{N}k\right],
\end{align}
where $\mathbf{L}^{(\mathrm{r})}$ and $\mathbf{L}^{(\mathrm{p})}$ denote the graph Laplacian matrices of the ring and path graphs, respectively. 
\subsection{Directed GFTs}
This section reviews conventional GFB/GFT methods for directed graphs \cite{MagGFT, AuGFT, SVDGFT, PDGFT, PDfilter, GDV, DGFT, HRWGFT, GenBC, StableApp, DiLap, DAGs}. These methods are classified into three categories: \textit{Eigenvalue-Decomposition-Based GFBs} (EVD-based GFBs) \cite{MagGFT, AuGFT, HRWGFT, DiLap}, \textit{Optimization-Based GFBs} (Opt-based GFBs) \cite{GDV, DGFT}, and other methods \cite{PDGFT, PDfilter, GenBC, StableApp, DAGs}.

\subsubsection{Eigenvalue-Decomposition-Based GFBs}\label{subsubsec:EVDGFB}
Unlike GFBs for undirected graphs, designing directed GFBs is challenging because adjacency matrices of directed graphs are not always diagonalizable. To address this, some conventional methods modify adjacency matrices so that the resulting graph Laplacian becomes diagonalizable.

In this section, we review a method utilizing the complex-valued graph Laplacian, known as the magnetic Laplacian, which encodes the directionality and connectivity of directed graphs \cite{MagGFT}. Specifically, for a given adjacency matrix $\mathbf{W} \in \mathbb{R}_+^{N \times N}$, the connectivity matrix is defined as $\mathbf{W}^{(s)} = \mathrm{Sym}(\mathbf{W}) \in \mathbb{R}_+^{N \times N}$, and the directionality matrix is given by $\mathbf{\Gamma}^{(q)} \in \mathbb{C}^{N \times N}$, where $\gamma^{(q)}_{n_1,n_2} = e^{\mathrm{j} 2\pi q (w_{n_1,n_2} - w_{n_2,n_1})}$ ($q \in [0,1)$ is referred to as the rotation parameter). The degree matrix is then defined as $[\mathbf{D}^{(s)}]_{n_1,n_1} = \sum_{n_2=1}^{N} w_{n_1,n_2}^{(s)}$, and the (Hermitian) magnetic Laplacian $\mathbf{L}^{(q)}  = \mathbf{L}^{(q)\mathrm{H}}\in \mathbb{C}^{N \times N} $ is given by:
\begin{align}\label{eq:maggfb}
\mathbf{L}^{(q)} := \mathbf{D}^{(s)} - \mathbf{\Gamma}^{(q)} \odot \mathbf{W}^{(s)} =  \mathbf{U}^{(q)}\mathbf{\Lambda}^{(q)}\mathbf{U}^{(q)\mathrm{H}},
\end{align}
where the eigenvectors $\mathbf{U}^{(q)} = [\mathbf{u}_{k}^{(q)}]_{k=1}^N\in \mathbb{C}^{N \times N}$ form a unitary matrix (referred to as \textit{MagGFB} in this paper), and the graph frequencies given by the eigenvalues are real, i.e., $\mathbf{\Lambda}^{(q)} = \mathrm{Diag}(\lambda_1^{(q)},\ldots,\lambda_N^{(q)} )\in \mathbb{R}^{N \times N}$. Consequently, the graph Fourier coefficients $\widehat{\mathbf{s}} \in \mathbb{C}^{N}$ can be obtained as $\widehat{\mathbf{s}} = \mathbf{U}^{(q)\mathrm{H}}\mathbf{s}$. For small values of the rotation parameter $q$, the magnetic Laplacian exhibits properties analogous to those of the ordinary graph Laplacian. Note that for undirected graphs, the magnetic Laplacian reduces to the ordinary graph Laplacian, i.e., $\mathbf{L}^{(q)} = \mathbf{L}$, and thus, the derived GFBs are identical.

\subsubsection{Optimization-Based GFBs}\label{subsubsec:OptGFB}
Another approach to designing directed GFBs involves minimizing the total variation of the directed graph \cite{GDV, DGFT}. This section describes the GFB with spread graph frequencies (SfGFB), which serves as the foundation for the optimization-based DGFF proposed in this study.

Graph frequencies are not always uniformly distributed over the graph spectral domain, as shown in Fig. \ref{fig:recim1}. A GFT with severely non-uniform graph frequencies results in biased spectral analysis of graph signals. To address this issue, the SfGFB was proposed in \cite{DGFT}. This method derives an orthogonal basis for the directed GFB $(\mathbf{U}_{\mathrm{Sf}} \in \mathbb{R}^{N\times N} )$ by minimizing a spectral dispersion function ${\delta(\mathbf{U})}: \mathbb{R}^{N\times N} \rightarrow \mathbb{R}_+$, which is based on the DV of directed graph signals $\mathcal{DV}: \mathbb{R}^{N} \rightarrow \mathbb{R}_+$ as follows:
\begin{align}\label{eq:DV}
\mathcal{DV}(\mathbf{x}) :=&\ \sum_{n_1,n_2=1}^{N} {w}_{n_1,n_2} [x_{n_1} - x_{n_2}]_+^2, \nonumber\\
\mathbf{U}_\mathrm{Sf} =&\ \argmin_{\mathbf{U}^\top \mathbf{U} = \mathbf{I} }\delta(\mathbf{U}) := \sum\limits_{k=1}^N [\mathcal{DV}(\mathbf{u}_{k+1}) - \mathcal{DV}(\mathbf{u}_k)]^2, \nonumber\\
\mathrm{s.t.}& \  \mathbf{u}_1 = \mathbf{u}_{\mathrm{min}} = \frac{1}{\sqrt{N}} \mathbf{1}_N,\  \mathbf{u}_N = \mathbf{u}_\mathrm{max},
\end{align}
where $\mathbf{u}_{\mathrm{min}}$ and $\mathbf{u}_{\mathrm{max}}$ are the vectors with minimal variation (DC component) and maximal variation, respectively.

The resulting GFB successfully achieves a uniform graph frequency distribution (a design example will be shown in Sec. \ref{sec:experimental}).
Note that, for undirected graphs, $\mathcal{DV}(\mathbf{x})$ is equivalent to $\mathcal{GTV}(\mathbf{x})$, making \eqref{eq:DV} applicable to undirected graphs as well.

\subsubsection{Other Directed GFBs}\label{subsubsec:GFFDict}
Other types of directed GFBs are based on matrix decomposition. One notable approach introduces a framework using polar decomposition (PD) to analyze directed graphs \cite{PDGFT, PDfilter}. By defining two complementary variations—indirect and in-flow—associated with the unitary and positive semidefinite components of the adjacency matrix, it provides deeper insights into graph signal behavior. This approach also includes computationally efficient graph filters that bypass costly spectral filtering.

Another key development involves SVD-based GFTs \cite{SVDGFT,SVDProduct}, which utilize singular value decompositions of the graph Laplacian to define graph frequencies and efficiently represent signals. These methods ensure numerical stability, align with classical GFT in undirected graphs, and are applicable to Cartesian product graphs \cite{SVDProduct}, effectively analyzing spatiotemporal data and outperforming magnetic Laplacian-based approaches in tasks such as denoising.

This work focuses on the DGFF design method for analytic-, EVD-, and Opt-based GFBs (see Table \ref{tab:compCSHT}), while the development of PD/SVD-based GFBs is left for future work.

\subsection{Graph Fourier/Wavelet Frames and Dictionaries}
The \textit{Redundant Graph Fourier Transform} (RGFT) has been proposed for undirected graphs, forming a Parseval frame \cite{RGFT}. The RGFT frame $\{\mathbf{f}_{k}\}_{k=1}^{2N} \subset \mathbb{R}^N$ consists of eigenvectors $\{\mathbf{u}_{k}\}_{k=1}^{N}  \subset \mathbb{R}^N$ from the original graph Laplacian $\mathbf{L} \in \mathbb{R}^{N \times N}$ and additional eigenvectors $\{\mathbf{v}_{k}\}_{k=1}^{N} \subset \mathbb{R}^N$ from the modified graph Laplacian $\mathbf{L} - 2\rho\mathbf{I}$. The parameter $\rho$ ensures the additional eigenvalues $\{\xi_k := \lambda_k + 2\rho\}_{k=1}^{N}$ are inserted between the original eigenvalues $\{\lambda_k\}_{k=1}^{N}$ as $\lambda_k \leq \xi_k = \lambda_k + 2\rho < \lambda_{k+1}$\footnote{Note that $\{\xi_k\}_{k=1}^{N}$ do not correspond to the original graph Laplacian $\mathbf{L}$.}.

Graph wavelet frames (GWFs) \cite{Shuman2015, Behjat2016, Behjat2025} analyze local spectral structures using graph spectral filter banks. By integrating spatial and spectral information, GWFs facilitate multiscale analysis. To ensure unbiased spectral analysis, filter banks are designed for uniform subband partitioning while considering input signals and graph structure.

From an SR perspective, dictionary learning \cite{Yankelevsky2016, Thanou2014, Yankelevsky2019} has been explored, incorporating graph structures to learn optimal dictionaries and improve SR. These methods adapt better to data than the GFB.

However, overcomplete systems have a drawback: The RGFT’s additional vectors and graph dictionary vectors do not explicitly correspond to the original graph Laplacian frequencies, limiting their use in spectral analysis and tasks like graph filtering. In contrast, our study assigns explicit intermediate graph frequencies to the new vectors, preserving the GFT-based spectral analysis advantages while extending its capabilities. Moreover, GWFs struggle with non-uniform graph Laplacian eigenvalues, leading to biased frequency analysis, and require sufficient graph signals for uniform subband filter bank design. Our DGFFs ensure uniform spectral analysis, potentially enhancing GWF performance.

\section{Denser-Graph-Frequency Graph Fourier Frames}\label{sec:DGFF}
In this section, we propose the DGFFs to form a frame with a denser graph frequency distribution. The DGFF incorporates additional vectors corresponding to graph frequencies that lie between the initial graph frequencies and constructs a redundant set of vectors. The DGFF contributes not only to unbiased frequency-domain analysis, similar to the SfGFB, but also to more efficient SR thanks to its redundancy. We introduce three types of DGFFs; ADGFFs in Sec. \ref{sec:ADGFFs}, LiDGFFs in Sec. \ref{sec:LiDGFFs}, and OptDGFFs in Sec. \ref{sec:OptDGFFs}. They can cover various types of undirected and directed GFBs. 
\subsection{Analytic DGFF}\label{sec:ADGFFs}
\begin{figure}[t]
\centering
        \begin{minipage}[b]{0.48\linewidth}
			\centering
				\scalebox{0.25}{\includegraphics[keepaspectratio=true]{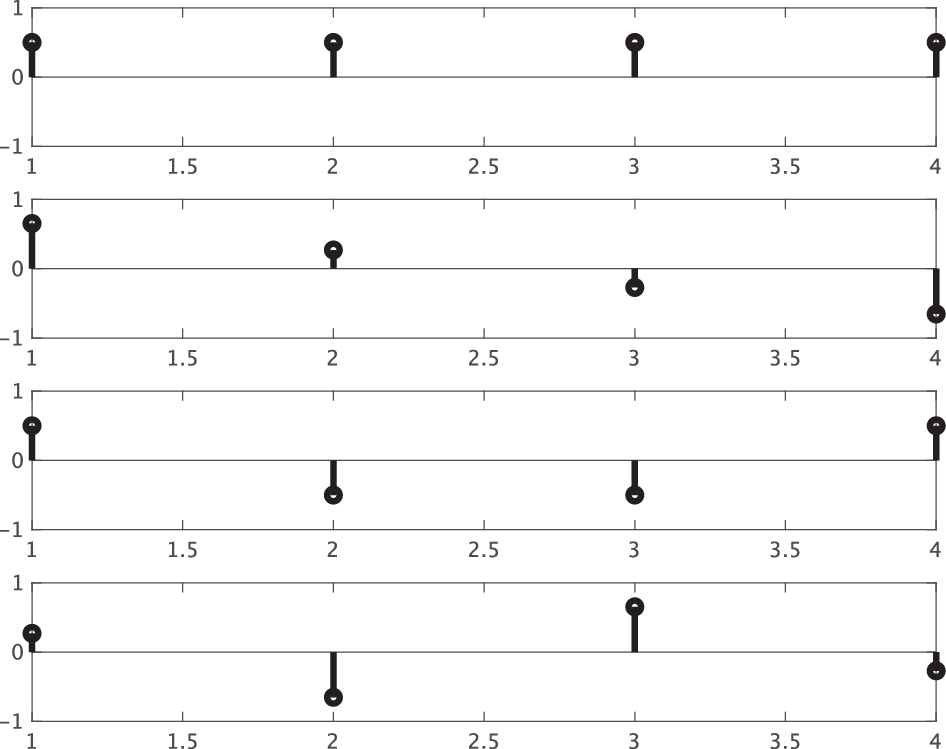}}
				\centerline{(a) Original vectors}
	\end{minipage}
 			\begin{minipage}[b]{0.48\linewidth}
			\centering
				\scalebox{0.25}{\includegraphics[keepaspectratio=true]{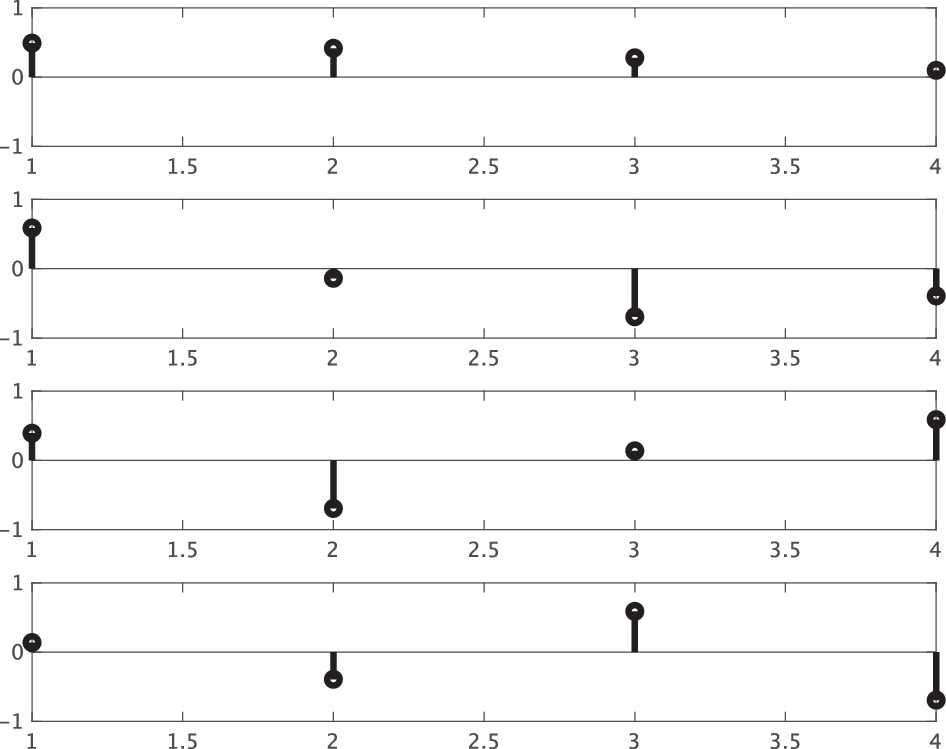}}
				\centerline{(b) Additional vectors}
	\end{minipage}
	\caption{
Original and additional oscillating vectors of $\mathbf{F}^{(\mathrm{p})}$ (Num. of nodes $N=4$).
 }\label{fig:dcb}
\end{figure}

This section introduces the ADGFFs. 
Since the GFBs for ring and path graphs can be defined as continuous (exponential and cosine) functions, we can easily generate the intermediate graph frequency vectors $\{\mathbf{u}_{k+1+\alpha}^{(\mathrm{r})}\}_{k=0}^{N-1} \subset \mathbb{C}^N$ and $\{\mathbf{u}_{k+1+\alpha}^{(\mathrm{p})}\}_{k=0}^{N-1} \subset \mathbb{R}^N$ ($\forall\alpha \in (0,1)$) as follows:
\begin{align}
[\mathbf{u}_{k+1+\alpha}^{(\mathrm{r})}]_\ell :=&\ {\frac{1}{\sqrt{N}}}\exp\left[\mathrm{j}\left(\frac{2\pi}{N}(k+\alpha)\right)\ell\right], \label{eq:ring_dgff}\\
[\mathbf{u}_{k+1+\alpha}^{(\mathrm{p})}]_\ell :=&\ c_k \sqrt{\frac{2}{N}} \cos\left[\left(\frac{\pi}{N}(k+\alpha)\right)\left(\ell+\frac{1}{2}\right)\right],\label{eq:path_dgff}
\end{align}
where $\alpha \in (0,1)$. Note that $\{\mathbf{u}_{k+1+\alpha}^{(\mathrm{r})}\}_{k=0}^{N-1}$ are unit vectors, while $\{\mathbf{u}_{k+1+\alpha}^{(\mathrm{p})}\}_{k=0}^{N-1}$ are not necessarily normalized. With this in mind, we define the ADGFFs for the ring and path as $\mathbf{F}^{(\mathrm{r})} = \begin{bmatrix} \mathbf{U}^{(\mathrm{r})} & [\mathbf{u}_{k +1+\alpha}^{(\mathrm{r})} ]_{k=0}^{N-1} \end{bmatrix} \mathbf{P}_1 \in \mathbb{C}^{N \times 2N}$ and $\mathbf{F}^{(\mathrm{p})} = \begin{bmatrix} \mathbf{U}^{(\mathrm{p})} & [\mathbf{u}_{k +1+\alpha}^{(\mathrm{p})}/\|\mathbf{u}_{k+1+\alpha}^{(\mathrm{p})}\|_2 ]_{k=0}^{N-1} \end{bmatrix}\mathbf{P}_1 \in \mathbb{R}^{N \times 2N}$, respectively, where the matrices $\mathbf{P}_1 \in \{0,1\}^{2N \times 2N}$ are permutation matrices arranging the vectors in graph frequency order. An example of the original and additional vectors for the path graph (the number of nodes: 4) is shown in Fig. \ref{fig:dcb}. Clearly, the additional vectors exhibit intermediate oscillation compared to the original GFB.

\subsection{Linear-Interpolation-Based DGFF}\label{sec:LiDGFFs}
Since eigenvectors are not necessarily represented by continuous functions, the ADGFF is not universally applicable to all undirected/directed graphs. In this section, we explain the LiDGFF which is one of two proposed methods for non analytic GFBs. The LiDGFF incorporates intermediate graph frequency vectors through the linear combination of neighboring graph frequency vectors. The graph frequencies of the derived vectors can be completely characterized based on the graph frequencies of the GFB obtained from the graph Laplacian or the normalized graph Laplacian on undirected graphs, as stated in the following theorem.
\begin{theorem}\label{theo1}
Let $\mathbf{L} \in \mathbb{C}^{N\times N}$ be an Hermitian matrix, $\mathbf{U} = \{\mathbf{u}_k\}_{k=1}^{N} \in \mathbb{C}^{N\times N}$  and $\{\lambda_k\}_{k=1}^{N} \subset \mathbb{R}$ be the eigenvectors and eigenvalues of $\mathbf{L}$, that is, $\mathbf{L} = \mathbf{U}\mathrm{Diag}(\lambda_1,\ldots,\lambda_{N}) \mathbf{U}^\mathrm{H}$. For $\alpha,\ \beta \in (0,1)$, let us define the intermediate graph frequency vectors $\{\mathbf{u}_{k}^{(\alpha,\beta)}\}_{k=1}^{N-1}$ as:
\begin{align}\label{IPDGFF}
\mathbf{u}_{k}^{(\alpha,\beta)} :=&\ \mathcal{IGV}(\mathbf{u}_{k},\mathbf{u}_{k+1},(\alpha,\beta)) \nonumber\\ :=&\ \frac{\alpha \mathbf{u}_{k} + \beta \mathbf{u}_{k+1}}{\left\|\alpha \mathbf{u}_{k} + \beta \mathbf{u}_{k+1}\right\|_2}.
\end{align}
Then, their graph frequencies are given by:
\begin{align}\label{IPDGFFfreq}
\mathbf{u}_{k}^{(\alpha,\beta)\mathrm{H}} \mathbf{L} \mathbf{u}_{k}^{(\alpha,\beta)}   = \frac{\alpha^2 }{\alpha^2 + \beta^2}\lambda_k + \frac{ \beta^2 }{\alpha^2 + \beta^2} \lambda_{k+1}.
\end{align}
\begin{proof}
From the orthogonal condition $\langle \mathbf{u}_k , \mathbf{u}_{k+1} \rangle = 0$, and the eigenvectors/eigenvalues $\{\mathbf{u}_k,\lambda_k\}$ satisfying $\mathbf{L}\mathbf{u}_k=\lambda_k\mathbf{u}_k$, we can derive as:
\begin{align}\label{eq:proof}
\left\|\alpha \mathbf{u}_k + \beta \mathbf{u}_{k+1}\right\|_2^2 =&\ \alpha^2 + \beta^2,\\
\mathbf{u}_k^\mathrm{H} \mathbf{L} \mathbf{u}_{k+1} =&\ \lambda_k \mathbf{u}_k^\mathrm{H} \mathbf{u}_{k+1} = 0.
\end{align}
Thus,
\begin{align}\label{eq:proof1}
\mathbf{u}_{k}^{(\alpha,\beta)\mathrm{H}} \mathbf{L} \mathbf{u}_{k}^{(\alpha,\beta)}   =&\ \frac{(\alpha \mathbf{u}_k + \beta \mathbf{u}_{k+1})^\mathrm{H}}{\left\|\alpha \mathbf{u}_k + \beta \mathbf{u}_{k+1}\right\|_2}\mathbf{L}\frac{\alpha \mathbf{u}_k + \beta \mathbf{u}_{k+1}}{\left\|\alpha \mathbf{u}_k + \beta \mathbf{u}_{k+1}\right\|_2}\nonumber\\
=&\ \frac{\alpha^2 \lambda_k + \beta^2\lambda_{k+1} }{\alpha^2 + \beta^2},
\end{align}
which completes the proof of \eqref{IPDGFFfreq}.
\end{proof}
\end{theorem}
Finally, the LiDGFF is defined as the union of the original GFB and the additional vectors $\mathbf{F} = \begin{bmatrix} \mathbf{U} & [\mathbf{u}_{k}^{(\alpha,\beta)}]_{k=1}^{N-1} \end{bmatrix}\mathbf{P}_2 \in \mathbb{C}^{N \times 2N-1}$, where $\mathbf{P}_2 \in \{0,1\}^{(2N-1) \times (2N-1)}$ is a permutation matrix arranging the vectors in graph frequency order. The intermediate graph frequency vectors and their corresponding frequencies of the GFB derived from the normalized graph Laplacian can also be defined by \eqref{IPDGFF} and \eqref{IPDGFFfreq}, with their frequencies forming a dense graph frequency distribution in the range of $(0, 2)$.
\begin{Rem}\upshape 
From Theorem \ref{theo1}, the LiDGFFs for undirected graphs can always be designed because the corresponding graph Laplacian is assumed to be symmetric. In addition, this method can be applicable to all the graphs with a diagonalizable graph Laplacian by a unitary matrix, which includes the directed ring graph. Even for general directed graphs, the MagGFB is designed from the Hermitian magnetic Laplacian $\mathbf{L}^{(q)}$, thus we can derive intermediate graph frequency vectors with graph frequencies defined in \eqref{IPDGFFfreq}\footnote{
MagGFB $\mathbf{U}^{(q)} = [\mathbf{u}_{k}^{(q)}]_{k=1}^N$ is a basis induced from the variation measure $\mathbf{x}^\mathrm{H} \mathbf{L}^{(q)} \mathbf{x}$. As described in \cite{MagGFT}, by setting the rotation parameter $q$ to a small value, the eigenvectors $\mathbf{u}_k^{(q)}$ are ordered in ascending order of eigenvalues such that they exhibit increasing total variation (TV), defined as:
\begin{align}
\mathcal{TV}(\mathbf{u}_k^{(q)}) = \sum_{(i,j) \in \mathcal{E}} \left| [\mathbf{u}_k^{(q)}]_i -[\mathbf{u}_k^{(q)}]_j \right|^2.
\end{align}
This ordering satisfies the relation  
$
\lambda_k^{(q)} \leq \lambda_{k+1}^{(q)} \Rightarrow \mathcal{TV}(\mathbf{u}_k^{(q)}) \leq \mathcal{TV}(\mathbf{u}_{k+1}^{(q)}).
$
Consequently, the intermediate graph frequency eigenvector of $\mathbf{u}_{k}^{(q)}$ and  $\mathbf{u}_{k}^{(q)}$ with $(\alpha, \beta)$, termed as $\mathbf{u}_{k}^{(q,\alpha,\beta)}$, corresponding to an eigenvalue satisfying  
$
\lambda_k^{(q)} \leq \lambda_{k}^{(q,\alpha,\beta)} \leq \lambda_{k+1}^{(q)},
$
also follows the total variation ordering:  
\begin{equation}
\mathcal{TV}(\mathbf{u}_k^{(q)}) \leq \mathcal{TV}(\mathbf{u}_{k+1}^{(q,\alpha, \beta)}) \leq \mathcal{TV}(\mathbf{u}_{k+1}^{(q)}).
\end{equation}
}.    
\end{Rem}
\subsubsection{Low-redundant LiDGFF}
The previously introduced DGFF methods generate redundant frames by inserting intermediate graph frequency vectors between all graph frequencies. However, depending on a graph structure, the resulting graph frequencies may already exhibit partially dense distributions or uniform spacing and do not need additional intermediate graph frequencies. In such scenarios, applying the LiDGFF directly can lead to overly redundant frames, increasing computational costs without substantially enhancing the uniformity of the graph frequency distribution. To reduce redundancy, we propose lrLiDGFF, which inserts intermediate graph frequency vectors only if the spacing between frequencies exceeds a given threshold. Specifically, for a given set of graph frequencies $\{\lambda_k\}_{k=1}^{N}$ $(0 \leq \lambda_k \leq \lambda_{k+1})$ and prespecified thresholding values $\{\mathcal{T}_{\ell}\}_{\ell=1}^L \subset \mathbb{R}_+ \cup \{+\infty\}$ ($\mathcal{T}_\ell < \mathcal{T}_{\ell+1}$), $\ell$ intermediate graph frequency vectors are generated and inserted only when $\mathcal{T}_\ell \leq {\lambda}_{k+1} - {\lambda}_k  < \mathcal{T}_{\ell+1}$ (see Algorithm \ref{alg:LrLiDGFF}). The intermediate graph frequency vectors are calculated for the LiDGFF as in \eqref{IPDGFF}. 

By arranging the generated intermediate graph frequency vectors in the appropriate graph frequency order, a low-redundancy frame can be constructed. As further discussed in Sec. \ref{sec:experimental}, lrLiDGFFs become particularly beneficial as the number of vertices increases, improving computational efficiency for signals on large graphs.

\begin{algorithm}[t]
    \caption{Low-redundant LiDGFFs}
    \label{alg:LrLiDGFF}
    \begin{algorithmic}[1]
        \STATE \textbf{Input:} The original GFB $\mathbf{U} = [\mathbf{u}_k]_{k=1}^{N}$, the original graph frequency $\mathbf{\Lambda} = \mathrm{Diag}(\lambda_1,\ldots,\lambda_{N})$, the weighting coefficients $\{(\alpha_{m}^{(\ell)},\beta_{m}^{(\ell)})\}_{\ell=1,\ldots,L,m = 1 , \ldots \ell}$, and the thresholding parameters $\{\mathcal{T}_\ell\}_{\ell=1}^L$, $\mathbf{U}^\star = \mathbf{U}$.
        \FOR{$k = 1$ to $N-1$}
        \FOR{$\ell = 1$ to $L-1$}
            \IF{$\mathcal{T}_\ell \leq \lambda_{k+1} - \lambda_{k} < \mathcal{T}_{\ell+1}$}
                \STATE Compute intermediate graph frequency vectors as:
                \[
                \mathbf{U}_{k}^{{(\ell)}} = [\mathcal{IGV}(\mathbf{u}_{k} , \mathbf{u}_{k+1}, (\alpha_{m}^{(\ell)},\beta_{m}^{(\ell)}))]_{m = 1}^{\ell}.
                \]
            \STATE Append the derived intermediate graph frequency vectors: $\mathbf{U}^\star \leftarrow \begin{bmatrix} 
            \mathbf{U}^\star & {\mathbf{U}_{k}^{{(\ell)}}}\end{bmatrix}$.
            \ENDIF
        \ENDFOR
        \ENDFOR
        \STATE \textbf{Output:} $\mathbf{U}^\star$.
    \end{algorithmic}
\end{algorithm}

\subsection{Optimization-based DGFF}\label{sec:OptDGFFs}
Next, we introduce the OptDGFFs for directed graphs, in the case where the GFB is not derived with the EVD-based approach, such as the MagGFB, but rather with the optimization-based one. The OptDGFFs can be applied to all directed GFBs that are derived by from differentiable graph total variations such as $\mathcal{GTV}(\mathbf{x})$ in \eqref{TV} or $\mathcal{DV}(\mathbf{x})$ in \eqref{eq:DV}. Here, we show the OptDGFF inspired by the SfGFB, which is termed as \textit{SfDGFF}. The SfGFB originally aims to achieve a more uniform compared to other GFB. Thus, by extending bases to frames, the proposed SfDGFF contributes not only to spreading graph frequency distribution but also to enhancing the efficiency of SR.

We explain the design of the SfDGFF in detail. 
First, we construct an optimization problem based on $\mathcal{DV}(\mathbf{x})$ to derive intermediate graph frequency vectors. We define an objective function  $\varphi_{\mathbf{U}}^{(\alpha,\beta)}:\mathbb{R}^{N \times (K-1)} \rightarrow \mathbb{R}_+$ with a matrix $\mathbf{U} = \begin{bmatrix}
    \mathbf{u}_1 & \cdots & \mathbf{u}_{K}
\end{bmatrix} \in \mathbb{R}^{N\times K}$ and $\alpha \in (0,1)$ and $\beta = 1-\alpha$ for deriving the intermediate graph frequency vectors  as follows:
\begin{align}\label{eq:DGFTDGFF1}
\varphi_{\mathbf{U}}^{(\alpha,\beta)}(\mathbf{X}) :=&\ \sum_{k=1}^{K-1} \alpha \bigl( \mathcal{DV}(\mathbf{u}_k) - \mathcal{DV}(\mathbf{x}_k)\bigr)^2 \nonumber\\
& \quad + \beta \bigl(\mathcal{DV}(\mathbf{u}_{k+1}) - \mathcal{DV}(\mathbf{x}_k)\bigr)^2 .
\end{align}
Based on the objective function and inspired from the design method of the SfGFB, we derive the intermediate graph frequency vectors $\mathbf{U}_{\mathrm{Sf}}^{(\alpha,\beta)} \in \mathbb{R}^{N \times (N-1)}$ for the SfDGFF as:
\begin{align}\label{eq:DGFTDGFF}
\mathbf{U}_{\mathrm{Sf}}^{(\alpha,\beta)} :=&\ \argmin_{\mathbf{X} \in{\mathbb{R}}^{N\times({N-1})},\ \mathbf{X}^{\top}\mathbf{X} = \mathbf{I}}\ \varphi_{\mathbf{U}_{\mathrm{Sf}}}^{(\alpha,\beta)}(\mathbf{X}),
\end{align}
where $\mathbf{U}_{\mathrm{Sf}} \in \mathbb{R}^{N\times N}$ is the SfGFB obtained from (\ref{eq:DV}). 
The SfDGFF is formulated as $\mathbf{F}_{\mathrm{Sf}} = \begin{bmatrix}
    \mathbf{U}_{\mathrm{Sf}} & \mathbf{U}_{\mathrm{Sf}}^{(\alpha,\beta)}
\end{bmatrix}\mathbf{P}_2$, ($\mathbf{P}_2 \in \{0,1\}^{(2N-1) \times (2N-1)}$ is a permutation matrix). The graph frequencies of the derived intermediate graph frequency vectors $\mathbf{U}_{\mathrm{Sf}}^{(\alpha,\beta)}$
can be calculated by ${\lambda}_{\mathrm{Sf}, k}^{(\alpha,\beta)} = \mathcal{DV}(\mathbf{u}_{\mathrm{Sf}, k}^{(\alpha,\beta)}) $. 

The problem in \eqref{eq:DGFTDGFF} is an optimization problem on the Stiefel manifold. In the field of GSP, optimization algorithms on the Stiefel manifold include methods such as the Cayley transform-based approach used in \cite{OptM} to solve (\ref{eq:DV}), as well as SOC and PAMAL employed in \cite{GDV} for deriving GFB bases \cite{SOC, PAMAL}. Additionally, other optimization methods such as those in \cite{SCF, TRDCM, PCAL} have also been proposed to solve problems on the Stiefel manifold. Among these, we adopt the parallelizable approach called Parallelizable Column-wise Block Minimization for PLAM (PCAL) to efficiently solve optimization problems on the manifold \cite{PCAL}. Since the gradient of the objective function is required, we consider the gradients of the objective functions used in the construction of the optimized DGFF, namely (\ref{eq:DGFTDGFF1}). 
First, each column of the gradient $\mathbf{G} = \begin{bmatrix} \mathbf{g}_1 & \cdots & \mathbf{g}_{K-1} \end{bmatrix} := \nabla\varphi_{\mathbf{U}}^{(\alpha,\beta)}(\mathbf{X})$ of the objective function $\varphi_{\mathbf{U}}^{(\alpha,\beta)}(\mathbf{X})$ in (\ref{eq:DGFTDGFF1}) is given as follows:
\begin{align}\label{DGFTDGFFgrad}
\mathbf{g}_k =& -2\frac{\partial}{\partial \mathbf{x}_k}\mathcal{DV}(\mathbf{x}_k) 
\Bigl(  \alpha \bigl(\mathcal{DV}(\mathbf{u}_k)  - \mathcal{DV}(\mathbf{x}_k) \bigr) \nonumber\\
&+  \beta \bigl(\mathcal{DV}(\mathbf{u}_{k+1}) -\nobreak\mathcal{DV}(\mathbf{x}_k) \bigr) \Bigr),
\end{align}
where the $n_1$-th element of the gradient of the directed variation is calculated as $\left[\frac{\partial}{\partial \mathbf{x}_k}\mathcal{DV}(\mathbf{x}_k) \right]_{n_1} = 2({\mathbf{W}}_{\cdot {n_1}}^{\top} [\mathbf{x}_k - {x_{ k,{n_1} }}\mathbf{1}_N]_+ - {\mathbf{W}}_{{n_1} \cdot} [{x_{ k,{n_1} }}\mathbf{1}_N - \mathbf{x}_k]_+ )$ \cite{DGFT}.
By solving the orthogonality-constrained optimization problem (\ref{eq:DGFTDGFF}) using PCAL with the gradient of the objective function $\mathbf{G}$, it is possible to generate a frame for graph frequency domain analysis based on the SfGFB. Finally, the algorithm for designing the SfDGFF by PCAL can be established as in Algorithm \ref{alg:OptDGFF}.

It should be noted that, according to \cite{PCAL}, PCAL requires the size of the variable $\mathbf{X} \in \mathbb{R}^{N \times K}$ with $N \geq 2K$. To satisfy this constraint, for a given SfGFB matrix $\mathbf{U}_{\mathrm{Sf}} = \begin{bmatrix}
    \mathbf{u}_{\mathrm{Sf},1} & \cdots &  \mathbf{u}_{\mathrm{Sf},N}
\end{bmatrix} \in \mathbb{R}^{N\times N}$, we set $\mathbf{U}_{\mathrm{Sf},1}$ and $\mathbf{U}_{\mathrm{Sf},2}$ containing the columns from the first to the $ \lceil \frac{N+1}{2} \rceil $-th column and from the $ \lceil \frac{N+1}{2} \rceil $-th to the $ N $-th column of the SfGFB matrix $\mathbf{U}_{\mathrm{Sf}}$, respectively. Based on this objective function, we then derive the intermediate graph frequency vectors $\mathbf{U}_{\mathrm{Sf},1}^{(\alpha,\beta)} \in \mathbb{R}^{N \times  (\lceil \frac{N+1}{2} \rceil -1)}$, $\mathbf{U}_{\mathrm{Sf},2}^{(\alpha,\beta)} \in \mathbb{R}^{N \times  (\lceil \frac{N+1}{2} \rceil -1  - \widetilde{i})} $ ($\widetilde{i}=0$ for odd $N$ and $\widetilde{i}=1$ for even $N$) by solving the optimization as: 
\begin{align}\label{eq:DGFTDGFF}
\mathbf{U}_{\mathrm{Sf},\widetilde{j}}^{(\alpha,\beta)} :=&\ \argmin_{ \mathbf{X}^{\top}\mathbf{X} = \mathbf{I}}\ \varphi_{\mathbf{U}_{\mathrm{Sf}, \widetilde{j}}}^{(\alpha,\beta)}(\mathbf{X}) \ (\widetilde{j}=1,2).
\end{align}
Note that the since the optimization problem is non-convex, it is not theoretically guaranteed that the graph frequencies of the derived vectors $\mathbf{U}_{\mathrm{Sf}}^{(\alpha,\beta)}$ are not guaranteed as $\mathcal{DV}(\mathbf{u}_{\mathrm{Sf},k}^{(\alpha,\beta)}) = \alpha\mathcal{DV}(\mathbf{u}_k) +  \beta\mathcal{DV}(\mathbf{u}_{k+1})$. In the experiments in Sec. \ref{sec:experimental}, we show the design example of the SfDGFF and evaluate error between ideal intermediate graph frequencies and those of the derived SfDGFF.

\begin{algorithm}[t]
    \caption{Solver of PCAL}
    \label{alg:OptDGFF}
    \begin{algorithmic}[1]
    \STATE \textbf{Input:} Given an initial estimate $\mathbf{X}^{[0]} = \begin{bmatrix} \mathbf{x}_1^{[0]} & \ldots & \mathbf{x}_{K}^{[0]} \end{bmatrix} \in \mathbb{R}^{N\times K}$ $(N \geq 2K)$, set the penalty parameter $\mu$ and the proximal parameter $\eta$. Initialize the iteration index as $i = 0$.
	\WHILE{certain stopping criterion is not reached}
		\STATE Compute the Lagrangian multipliers:
                        \begin{flalign*} 
                \mathbf{\Lambda}^{[i]} &= \mathrm{Sym}(\nabla \varphi_{\mathbf{U}}^{(\alpha,\beta)}(\mathbf{X}^{[i]})^{\top}\mathbf{X}^{[i]}) \nonumber\\ 
                &+ \mathrm{Diag}(\mathrm{diag}\bigl({\mathbf{X}^{[i]}}^{\top}\times\\&\nabla_{\mathbf{X}} \mathcal{L}_{\mu}(\mathbf{X}^{[i]}, \mathrm{Sym}(\nabla \varphi_{\mathbf{U}}^{(\alpha,\beta)}(\mathbf{X}^{[i]})^{\top}\mathbf{X}^{[i]}))\bigr) \nonumber
                \end{flalign*} 
	
		\FOR{$k = 1,...,K$}
			\STATE Minimize the following proximal linearized Lagrangian function:
                             \begin{flalign*} 
              \mathbf{x}_k^{[i+1]} & = \argmin_{\mathbf{x}\in\mathbb{R}^N} \widetilde{\mathcal{L}}_{\mu}^{(k)}(\mathbf{x}) \nonumber\\
            \widetilde{\mathcal{L}}_{\mu}^{(k)}(\mathbf{x})& := \nabla_{\mathbf{X}}\mathcal{L}_{\mu}(\mathbf{X}^{[i]}, \mathbf{\Lambda}^{[i]})_k^{\top}(\mathbf{x} - \mathbf{x}_k^{[i]})\\
            & + \frac{\eta^{[i]}}{2}\| \mathbf{x}-\mathbf{x}_k^{[i]} \|_2^2 \nonumber\\ 
            & \mathrm{s.t.}\  \| \mathbf{x} \|_2 = 1 
            \end{flalign*}
			\STATE Update $\mathbf{X}^{[i+1]}=[\mathbf{x}_{1}^{{[i+1]}},\dots,\mathbf{x}_{K}^{{[i+1]}}]$, \\
                       and set $i:=i+1$.
		\ENDFOR
	\ENDWHILE
	\STATE \textbf{Output:} $\mathbf{X}^{[i]}$.
    \end{algorithmic}
\end{algorithm}

\section{Denser-Graph-Spectral Filtering with Denser-Graph-Frequency Graph Fourier Frames}\label{sec:DGFFfiltering}
This section introduces the method for designing graph spectral filters using the DGFF, i.e., DGS filtering. Designing filters based on the DGFF presents a challenge, as it is difficult to directly obtain graph Fourier coefficients through matrix transformations, as is possible with conventional GFT, due to the overcomplete nature of DGFF matrices. To address this issue, this study employs an optimization problem based on the $\ell_1$ norm to derive sparse graph Fourier coefficients, enabling the design of graph spectral filters using DGFF.

For a given graph signal $\mathbf{s} \in \mathbb{R}^N$ and the designed DGFF $\mathbf{F}\in \mathbb{C}^{N\times K}$, the sparse coefficients $\widetilde{\mathbf{a}} \in \mathbb{C}^K$ used in DGS filtering can be obtained by solving the following optimization problem (detail information for its algorithm, see Appendix \ref{sec:AIR}): 
\begin{align}\label{eq:DGFFcoef} 
\widetilde{\mathbf{a}} = \mathcal{F}(\mathbf{s}) = 
\argmin_{\mathbf{a} \in \mathbb{C}^K} \|\mathbf{a}\|_1 \quad \mathrm{s.t.}\ \mathbf{F}\mathbf{a} = \mathbf{s}. 
\end{align} 
After obtaining these sparse coefficients, we define the filter frequency response $\mathbf{h}_\mathbf{\Lambda} = \begin{bmatrix}
    h(\lambda_1) & \cdots & h(\lambda_{K})
\end{bmatrix}^\top$, and subsequently, compute the filtered signal $\widetilde{\mathbf{s}}$ can be computed as follows: 
\begin{align}\label{eq:DGFFfilter}
\widetilde{\mathbf{s}} 
&=  \mathcal{F}^{-1}(\mathrm{Diag}(\mathbf{h}_{\boldsymbol{\Lambda}})\mathcal{F}(\mathbf{s}))
=  \mathbf{F}\mathrm{Diag}(\mathbf{h}_{\boldsymbol{\Lambda}}){\widetilde{\mathbf{a}}}.
\end{align}
We term the procedure in \eqref{eq:DGFFfilter} as DGS filtering. It enables finer extraction of target frequency components from input signals than GS filtering. We demonstrate its performance in the experiments in Sec. \ref{sec:experimental}.

\section{Experimental Results}\label{sec:experimental}
This section presents a comparative analysis of the proposed DGFF (LiDGFF/lrLiDGFF/SfDGFF) and the conventional GFBs in terms of denoising by GS filtering for undirected and directed graph signals, as well as graph signal recovery\cite{Naganuma2023} from subsampled signals using SR under the absence or presence of additive Gaussian noise. For all evaluations, the weighting parameters for the DGFFs were set to $(\alpha, \beta) = (0.5, 0.5)$. Additionally, the thresholding values for lrLiDGFF were set as $\mathcal{T}_1 = \mathcal{T}$ and $\mathcal{T}_2 = \infty$, where $\mathcal{T} := \frac{1}{3(N-1)} \sum_{k=1}^{N-1} (\lambda_{k+1} - \lambda_k)$. All experiments were implemented using MATLAB 2023a on MacOS with an Apple M2 chip (3.49GHz) and 8GB of RAM.

\subsection{Design Examples of DGFFs}
\begin{figure}[t] 
  \centering 
 	  \scalebox{0.3}{\includegraphics[keepaspectratio=true]{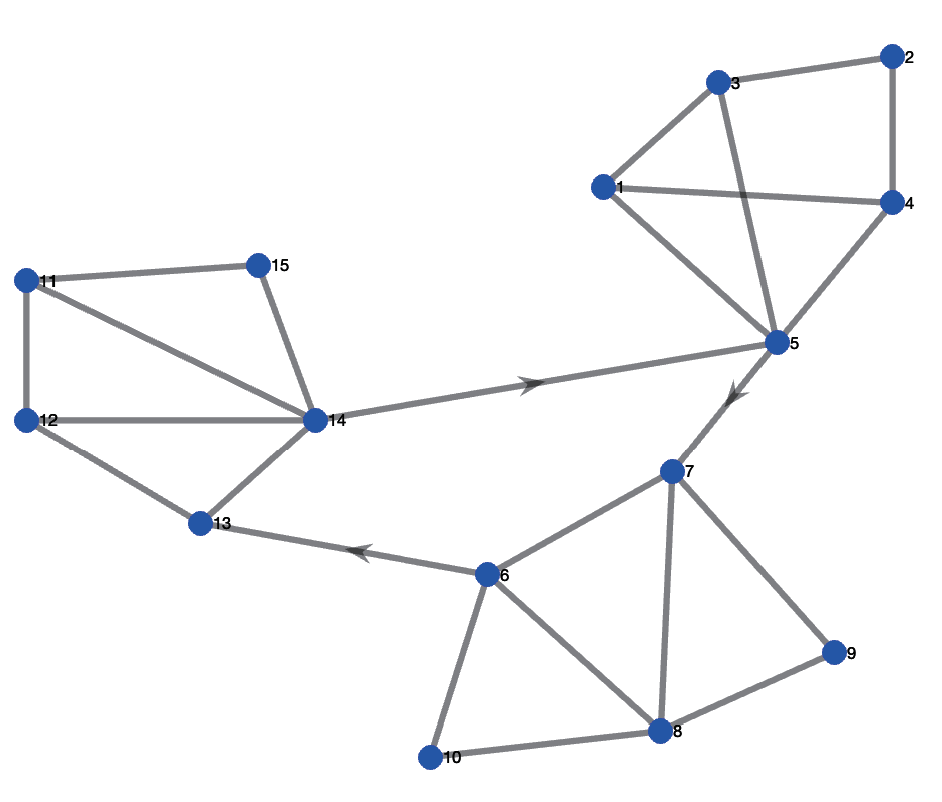}}
    \caption{Synthetic directed graph ($N=15$)} \label{fig:UDSG}
 \end{figure}

First, we consider an undirected graph with an adjacency matrix $\widehat{\mathbf{W}} \in \mathbb{R}^{15 \times 15}$ constructed by symmetrization $\widehat{\mathbf{W}} := \mathrm{Sym} (\mathbf{W}_{15})$ of an adjacency matrix $\mathbf{W}_{15}$ defined for a directed graph with 15 nodes (see Fig. \ref{fig:UDSG}) presented in \cite{GDV}, where the weights $[\mathbf{W}_{15}]_{n_1,n_2} = 1$ if the $n_1$-th node is connected to the $n_2$-th node and otherwise $[\mathbf{W}_{15}]_{n_1,n_2}  = 0$. Then we compare the graph frequency distributions of the conventional GFB and the proposed LiDGFF and lrLiDGFF derived from the GFB. 

\begin{figure}[t]
\centering
    \begin{minipage}[b]{0.48\linewidth}
    \centering
        \scalebox{0.1775}{\includegraphics[keepaspectratio=true]{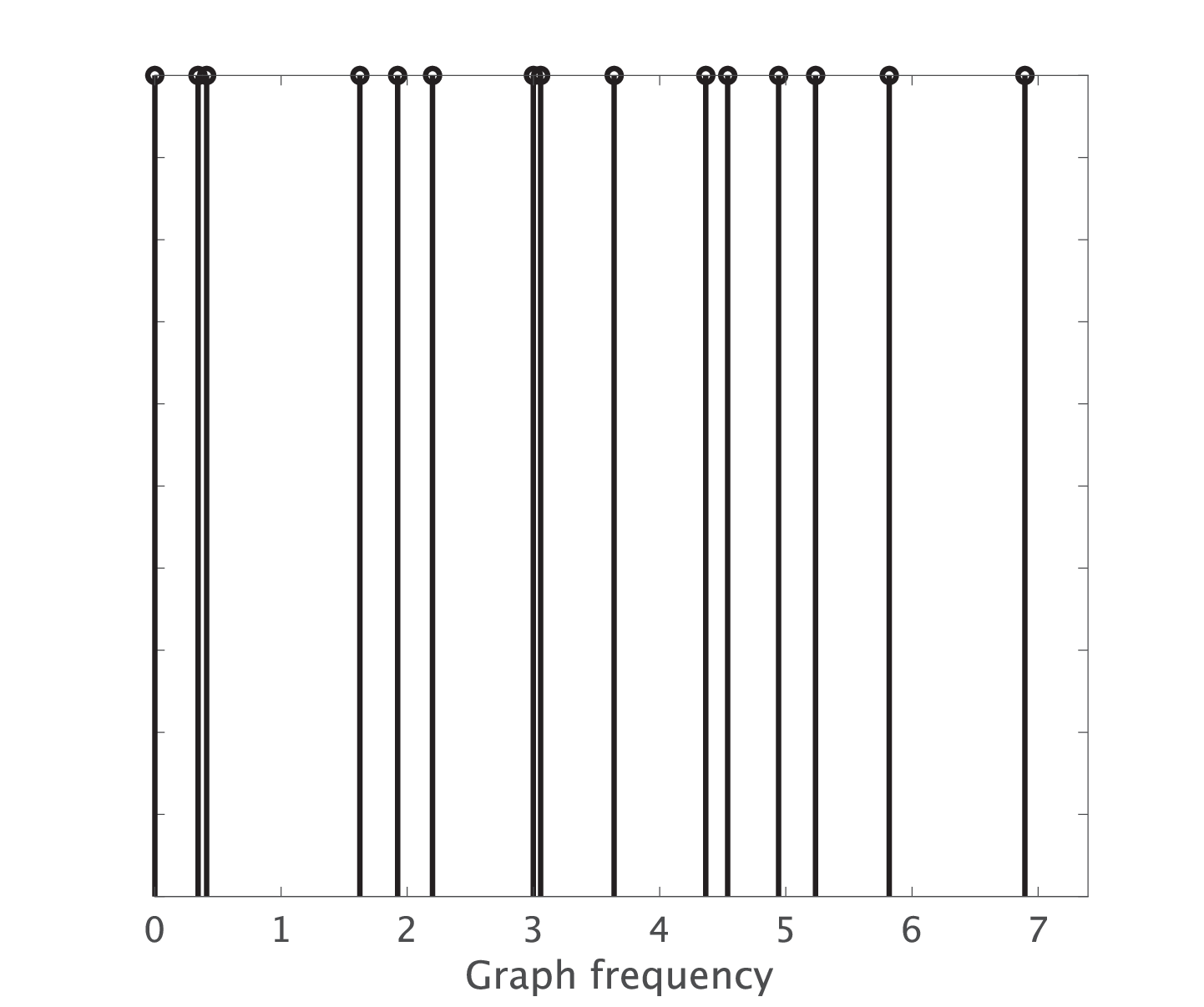}}
        \centerline{(a) GFB}
    \end{minipage}
    \begin{minipage}[b]{0.48\linewidth}
    \centering
        \scalebox{0.1775}{\includegraphics[keepaspectratio=true]{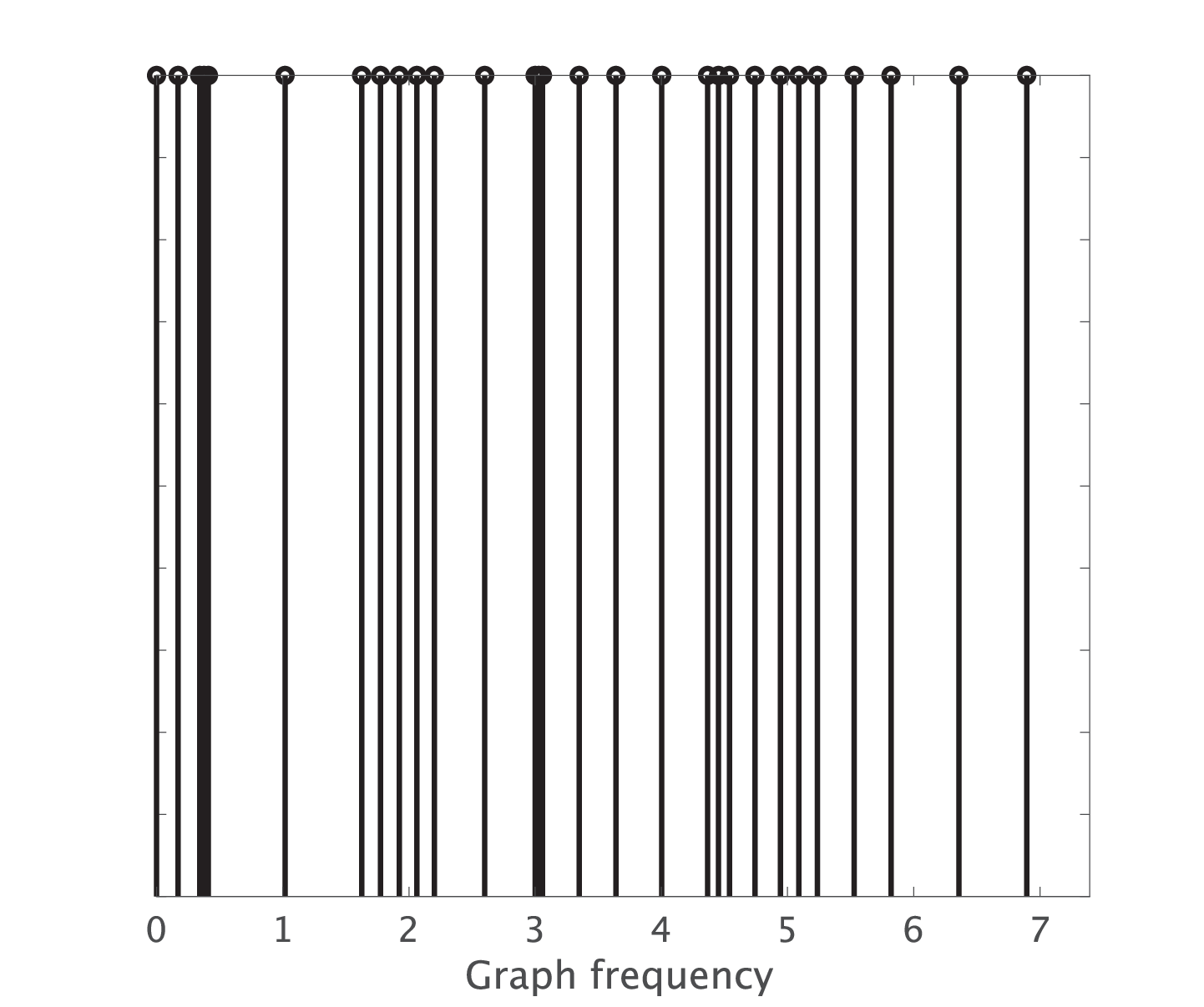}}
        \centerline{(b) LiDGFF}
    \end{minipage}
    \begin{minipage}[b]{0.48\linewidth}
    \centering
        \scalebox{0.1775}{\includegraphics[keepaspectratio=true]{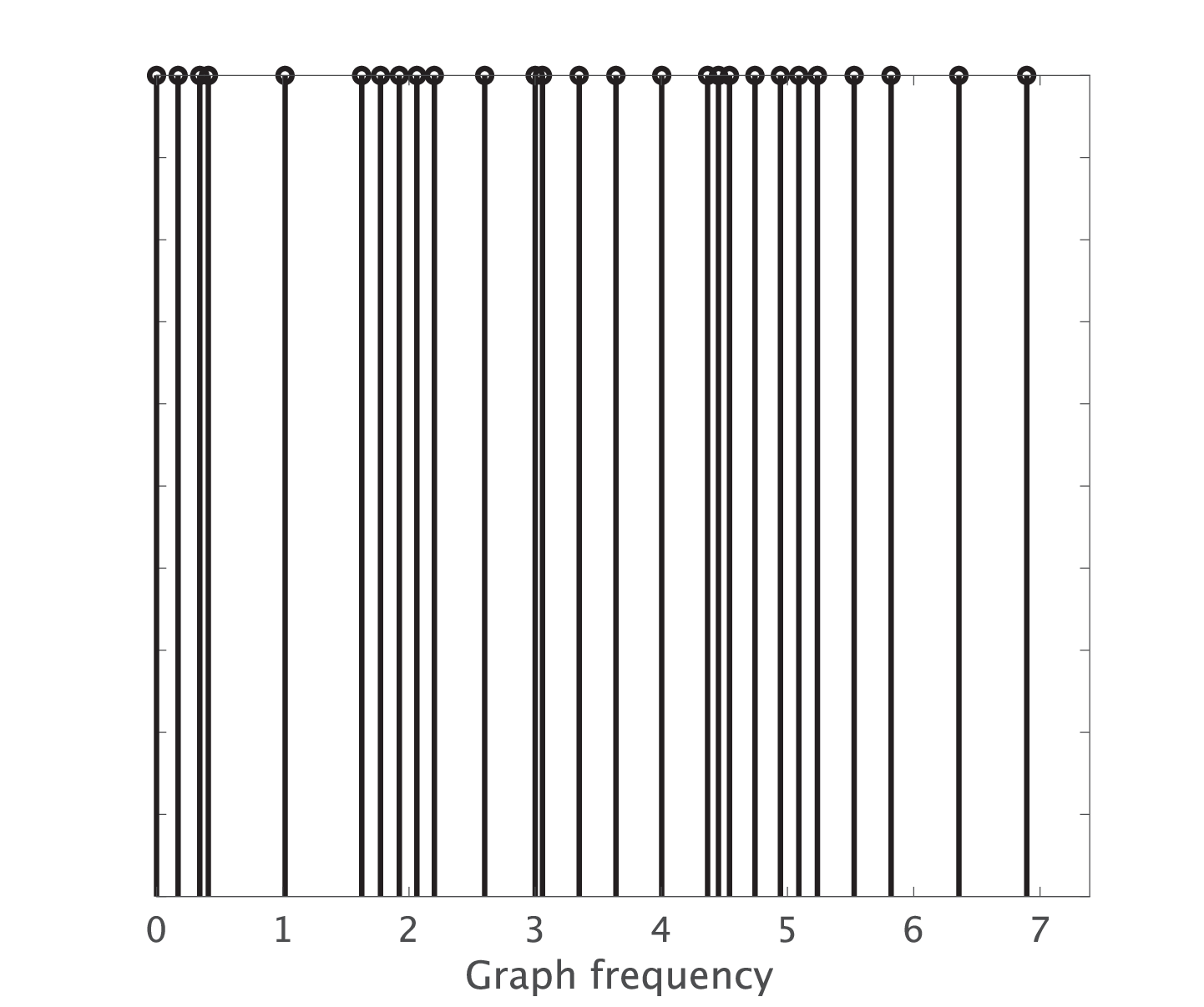}}
        \centerline{(c) lrLiDGFF ($\mathcal{T}_1$)}
    \end{minipage}
        \begin{minipage}[b]{0.48\linewidth}
    \centering
 	  \scalebox{0.1775}{\includegraphics[keepaspectratio=true]{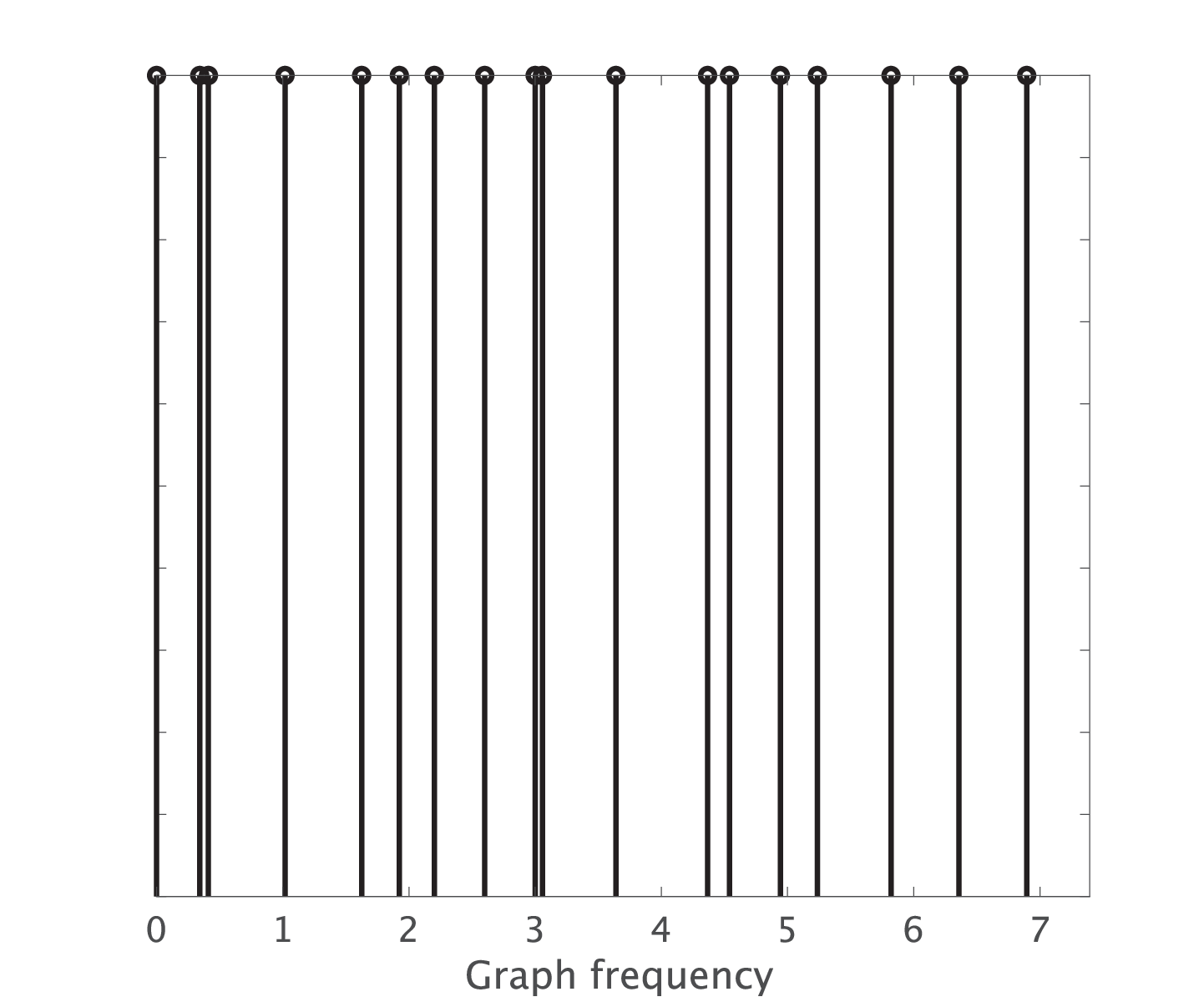}}
    \centerline{(d) lrLiDGFF ($\frac{9}{2}\mathcal{T}_1$)}
    \end{minipage}
\caption{Graph frequency distributions for GFB, LiDGFF and lrLiDGFFs, with thresholding values $\mathcal{T}_1$ and $\frac{9}{2} \mathcal{T}_1$.}
\vspace{-0.5cm}
\label{UDGfreq}
\end{figure}

The resulting graph frequency distributions are shown in Fig. \ref{UDGfreq}. As clearly illustrated in Fig. \ref{UDGfreq}(a), the conventional GFB exhibits a biased distribution. On the other hand, the LiDGFF with 29 frequency components in Fig. \ref{UDGfreq}(b) contains additional graph frequencies that are absent from the original GFB. Consequently, the LiDGFF increases the overall density of the graph frequency distribution. As another design example, the lrLiDGFFs with a threshold of $ \mathcal{T}_1$ and $\frac{9}{2}\mathcal{T}_1$ are shown in Figs. \ref{UDGfreq}(c) and (d). They reduce redundant intermediate graph frequency vectors while accounting for the spacing between frequencies. Next, these methods are numerically compared using the spectral dispersion $\delta(\cdot)$ defined in \eqref{eq:DV}. As shown in Table \ref{table:SDF}, the proposed DGFF achieves lower dispersion compared to both GFB. Additionally, even with the high threshold and minimal insertions in the lrLiDGFF, the dispersion remains lower than that of GFB.
In terms of the total number of inserted vectors, the lrLiDGFF reduces the count compared to the 29 vectors of LiDGFF, with 27 vectors ($\mathcal{T}_1$) and 21 vectors ($\frac{9}{2}\mathcal{T}_1$), respectively. 

\begin{figure}[t]
\centering
    \begin{minipage}[b]{0.48\linewidth}
    \centering
        \scalebox{0.1775}{\includegraphics[keepaspectratio=true]{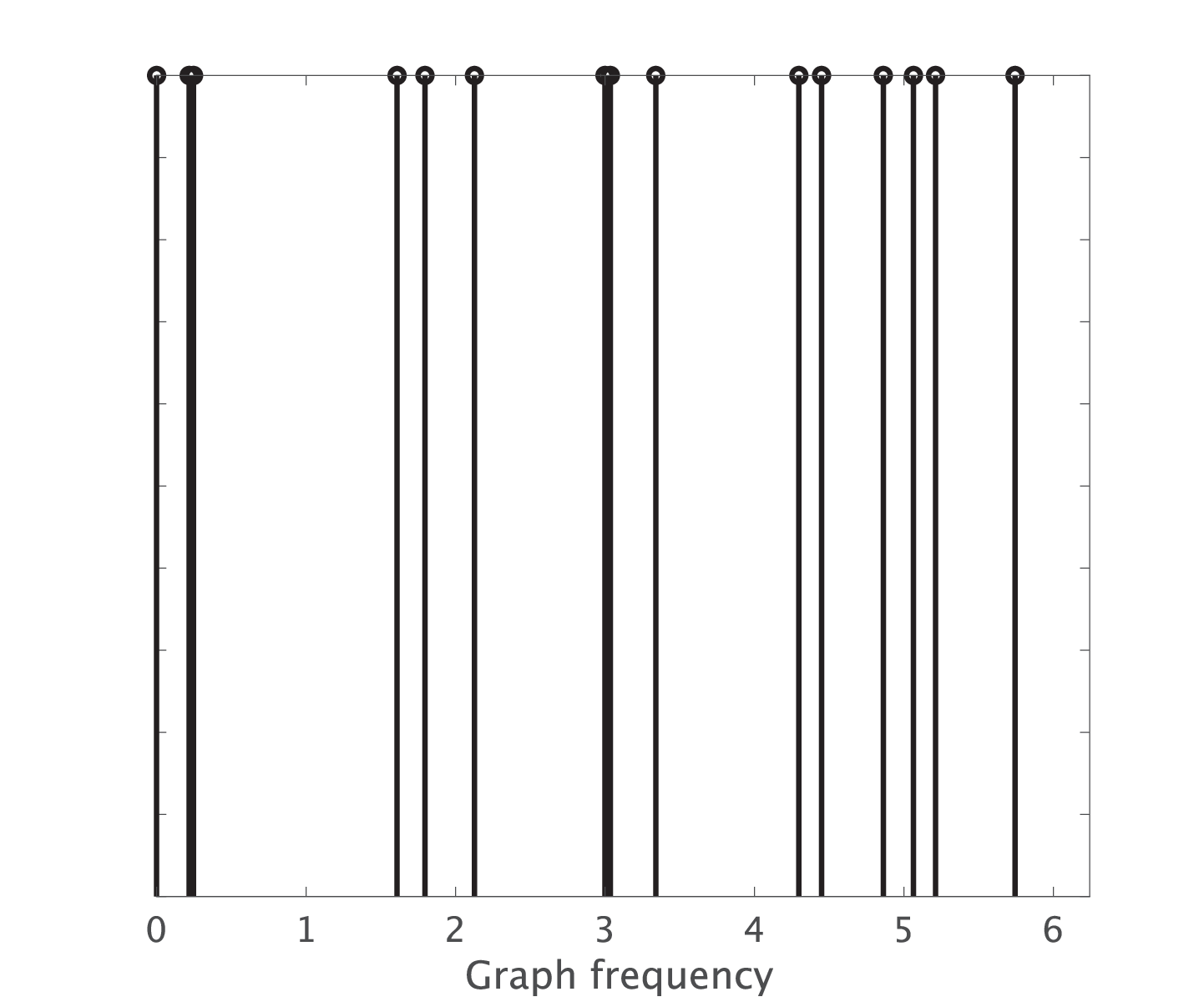}}
        \centerline{(a) MagGFB}
    \end{minipage}
    \begin{minipage}[b]{0.48\linewidth}
    \centering
        \scalebox{0.1775}{\includegraphics[keepaspectratio=true]{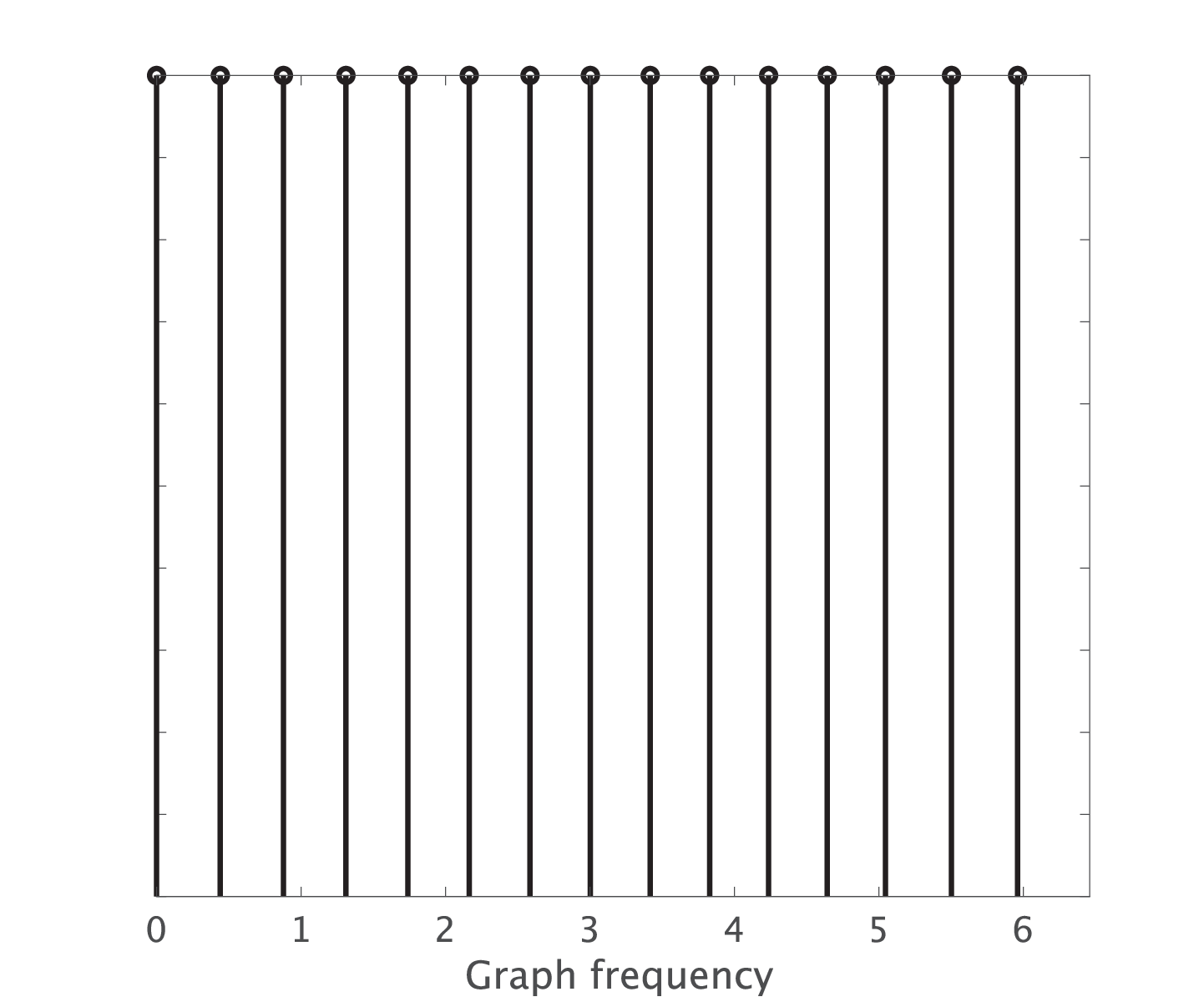}}
        \centerline{(b) SfGFB}
    \end{minipage}
    \begin{minipage}[b]{0.48\linewidth}
    \centering
        \scalebox{0.1775}{\includegraphics[keepaspectratio=true]{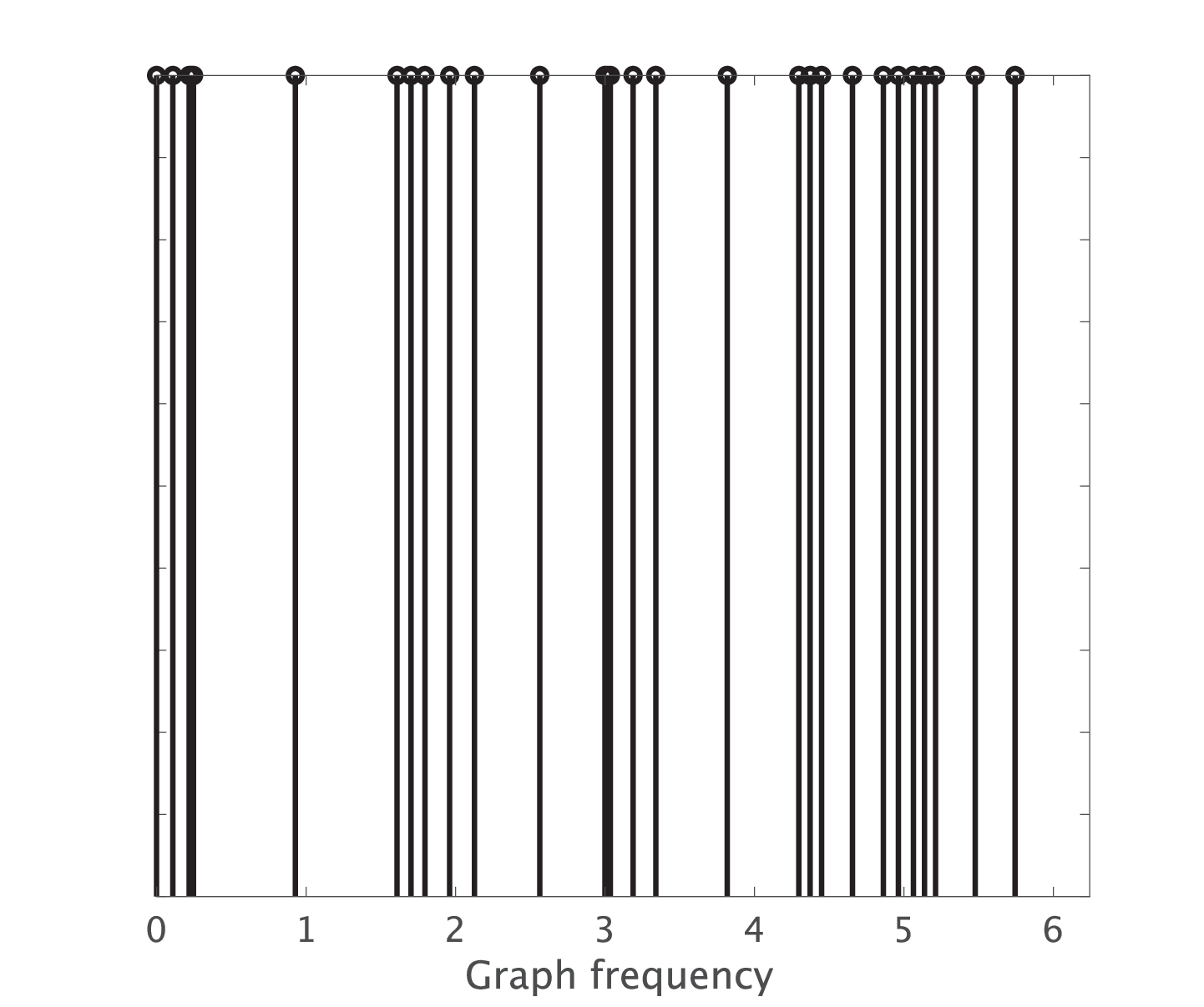}}
        \centerline{(c) MagDGFF}
    \end{minipage}
        \begin{minipage}[b]{0.48\linewidth}
    \centering
 	  \scalebox{0.1775}{\includegraphics[keepaspectratio=true]{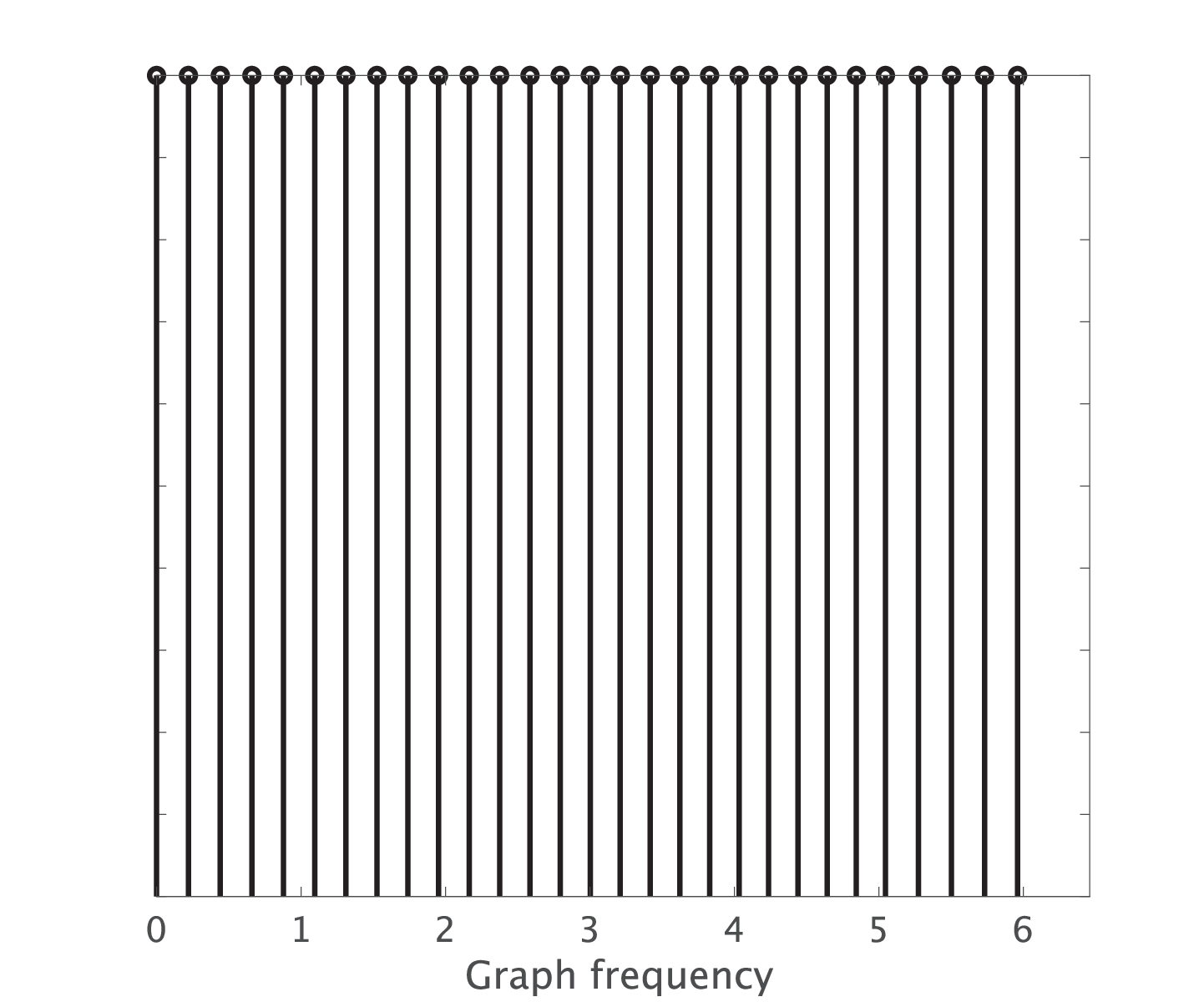}}
    \centerline{(d) SfDGFF}
    \end{minipage}
\caption{Graph frequency distributions of MagGFB, SfGFB, MagDGFF, and SfDGFF obtained from the directed graph (Fig. \ref{fig:UDSG})}
\label{DGfreq}
\end{figure}

Similarly, based on the adjacency matrix $\mathbf{W}_{15}$ of the directed graph, we compared the graph frequency distributions of MagGFB, SfGFB, the LiDGFF extension of MagGFB (MagDGFF), and SfDGFF. As shown in Fig. \ref{DGfreq}, in each case, the DGFF achieved a higher overall density of the graph frequency distribution compared to the conventional GFB. Notably, the SfDGFF maintained the uniformity of the graph frequency distribution inherent to SfGFB while achieving a higher density. Furthermore, the uniformity of the graph frequency distributions for directed graphs was evaluated using spectral dispersion in Table \ref{table:SDF}. As a result, each DGFF demonstrated lower dispersion compared to its corresponding GFB.

\begin{table}[t]
\caption{Spectral Dispersion of Different Algorithms for the Undirected/Directed Graph in Fig. \ref{fig:UDSG}}
\vspace{-0.2cm}
\label{table:SDF}
\begin{center}
\scalebox{1}{ 
\begin{tabular}{c|c|c|c} \thline
\multicolumn{2}{c|}{\mbox{Undirected Graph}} & \multicolumn{2}{c}{\mbox{Directed Graph}}\\ \thline
   {Method} & {Dispersion}  &    {Method} & {Dispersion}  \\ \hline
   GFB & 5.045 &  MagGFB & 4.358  \\ \hline
   {LiDGFF} & \textbf{2.522} & SfGFB & 2.541 \\ \hline
   {lrLiDGFF ($\mathcal{T}_1$)} & \textbf{2.526} & {MagDGFF} & \textbf{2.179}  \\ \hline
   {lrLiDGFF ($\tfrac{9}{2}\mathcal{T}_1$)} & \textbf{3.411} & {SfDGFF} & \textbf{1.270}  \\ \thline
 \end{tabular}
 }
 \end{center}
 \vspace{-0.3cm}
\end{table}

\begin{figure}[t]
    \centering
    \begin{minipage}{0.92\linewidth}  
        \centering
        \includegraphics[width=1.0\linewidth]{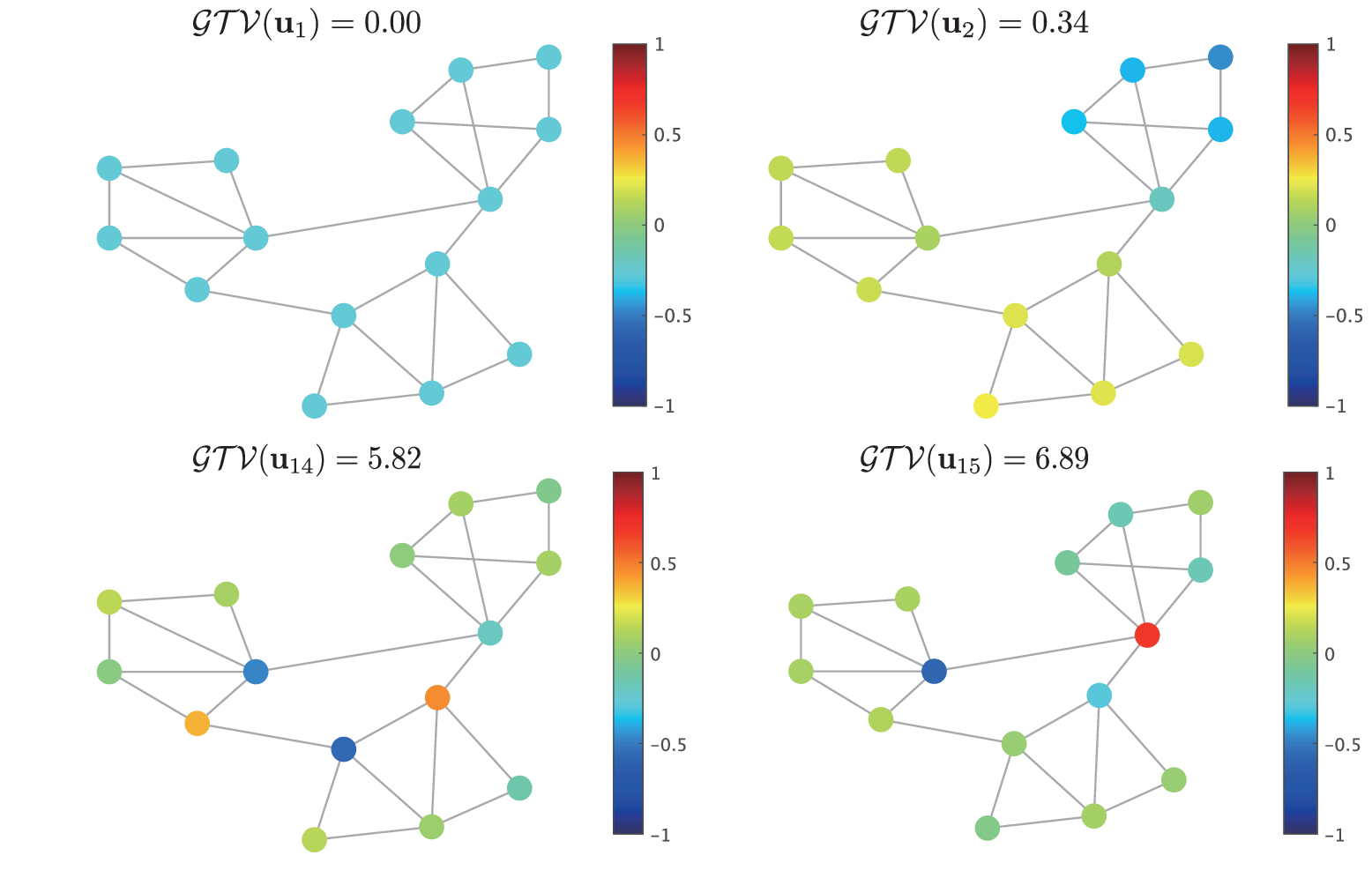}
        \caption{GFB vectors obtained from the synthetic undirected graph in Fig. \ref{fig:UDSG} and their corresponding graph frequencies.}
        \label{GFBplot}
    \end{minipage}

    \vspace{0.3cm} 

    \begin{minipage}{0.925\linewidth}  
        \centering
        \includegraphics[width=1.0\linewidth]{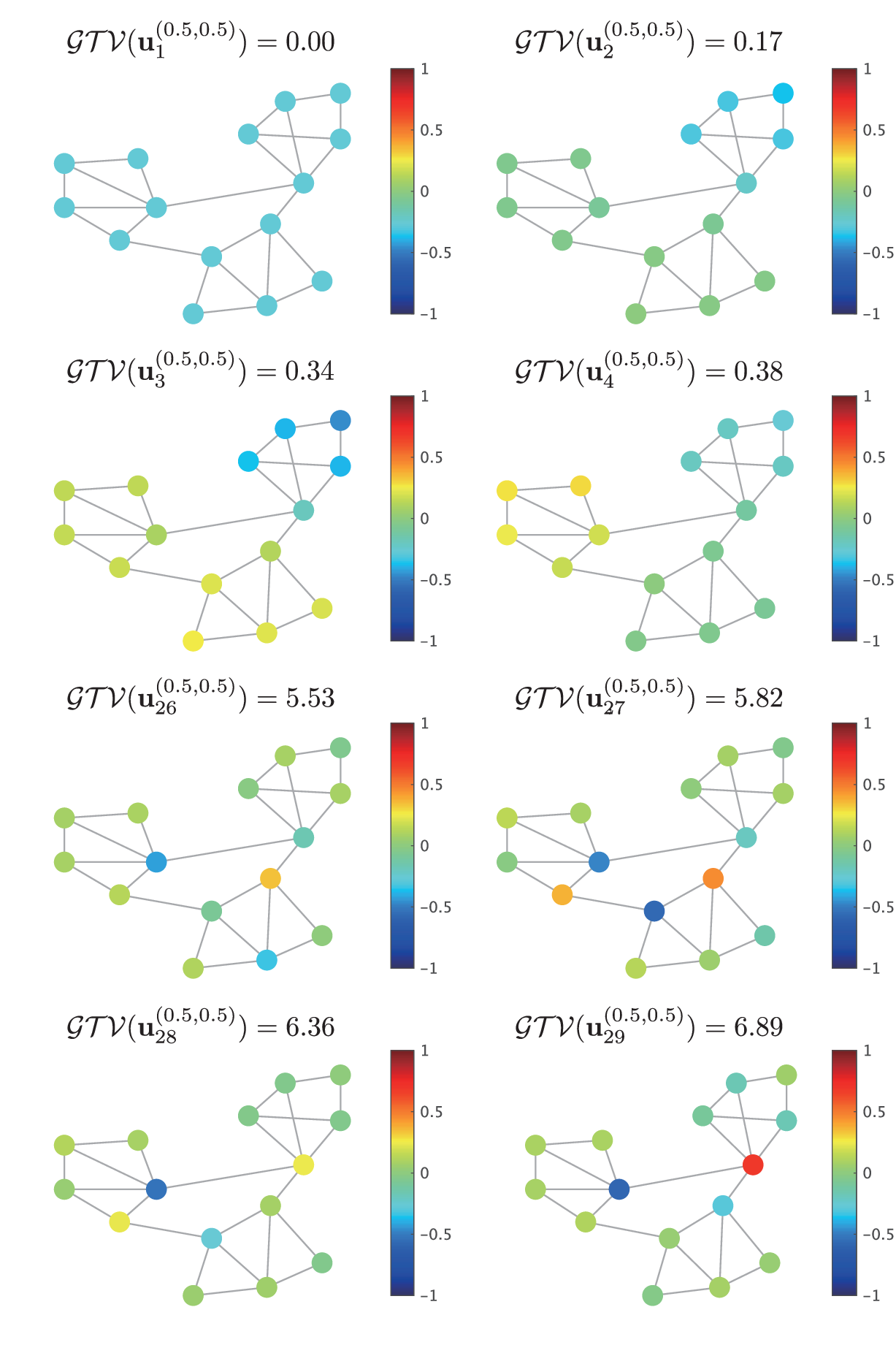}
        \caption{DGFF vectors obtained from the synthetic undirected graph in Fig. \ref{fig:UDSG} and their corresponding graph frequencies.}
        \label{DGFFplot}
    \end{minipage}
    \vspace{-0.5cm}  
\end{figure}

Figs. \ref{GFBplot} and \ref{DGFFplot} illustrate the frequency components of the GFB and the LiDGFF for the undirected graph derived from the directed one shown in Fig. \ref{fig:UDSG}. Both Figs. \ref{GFBplot} and \ref{DGFFplot} illustrate vectors with low frequencies at the top and higher frequencies towards the bottom. It can be observed that LiDGFF achieves a smoother transition between frequency components compared to the GFB by incorporating smoothly varying components as intermediate graph frequency vectors. 

\subsection{Graph Spectral Filtering}\label{sec:filtering}
\subsubsection{Simple GS and DGS Filtering}\label{subsec:SGFDGS}
\begin{figure}[t]
\centering
    \begin{minipage}[b]{0.49\linewidth}
    \centering
        \scalebox{0.19}{\includegraphics[keepaspectratio=true]{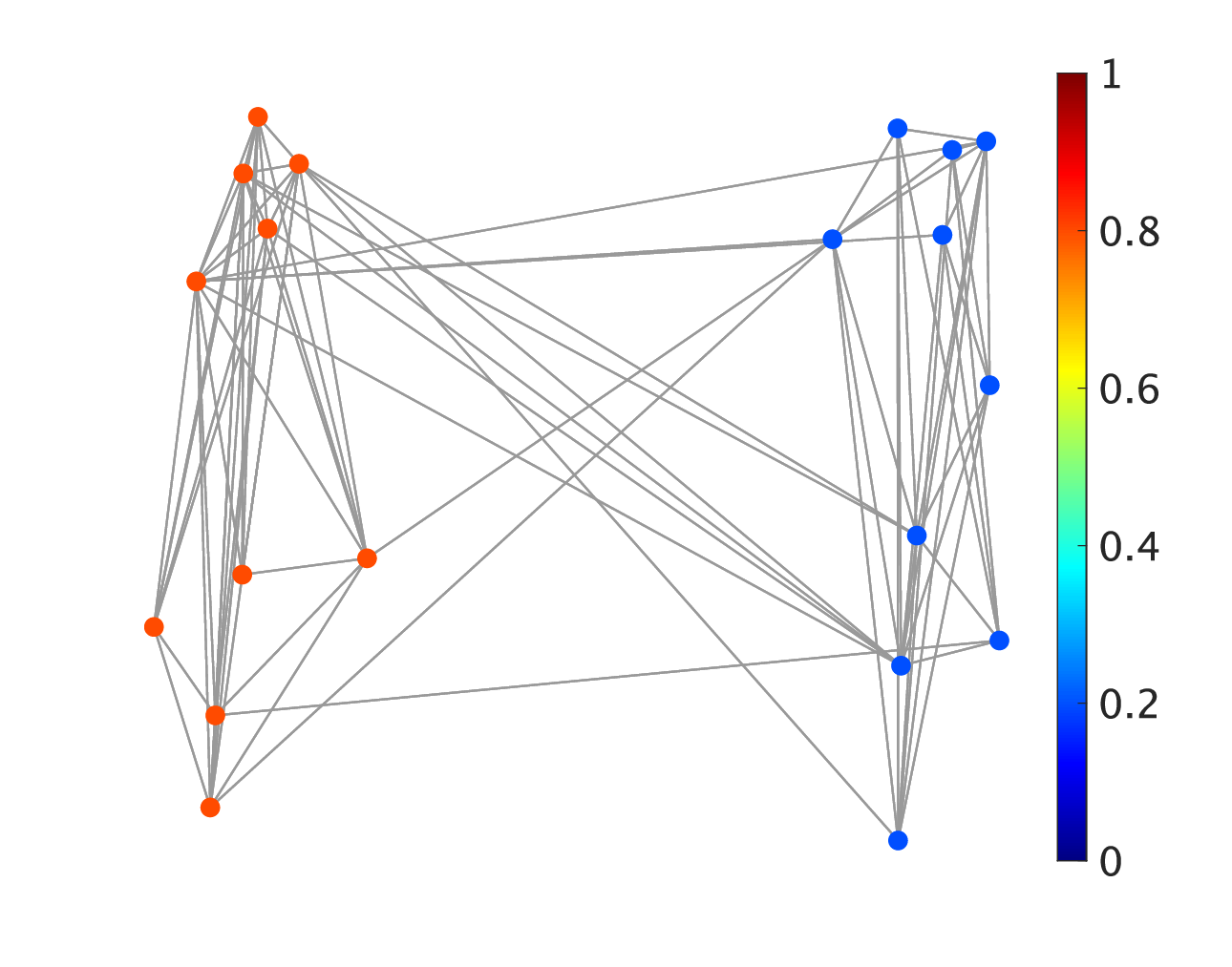}}
        \centerline{(a) Original signal}
    \end{minipage}
    \begin{minipage}[b]{0.49\linewidth}
    \centering
        \scalebox{0.19}{\includegraphics[keepaspectratio=true]{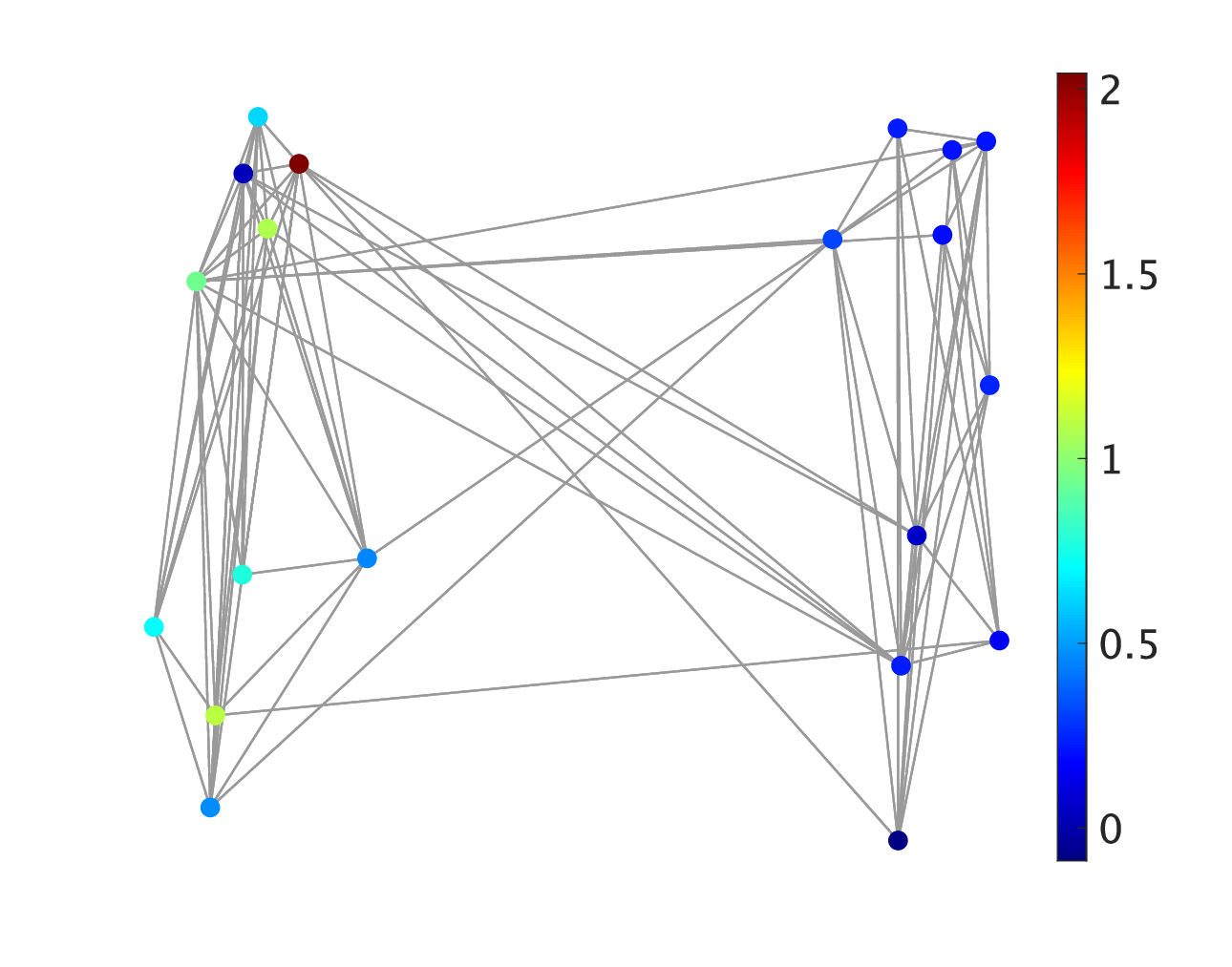}}
        \centerline{(b) Noise signal}
    \end{minipage}
        \begin{minipage}[b]{0.49\linewidth}
    \centering
        \scalebox{0.23}{\includegraphics[keepaspectratio=true]{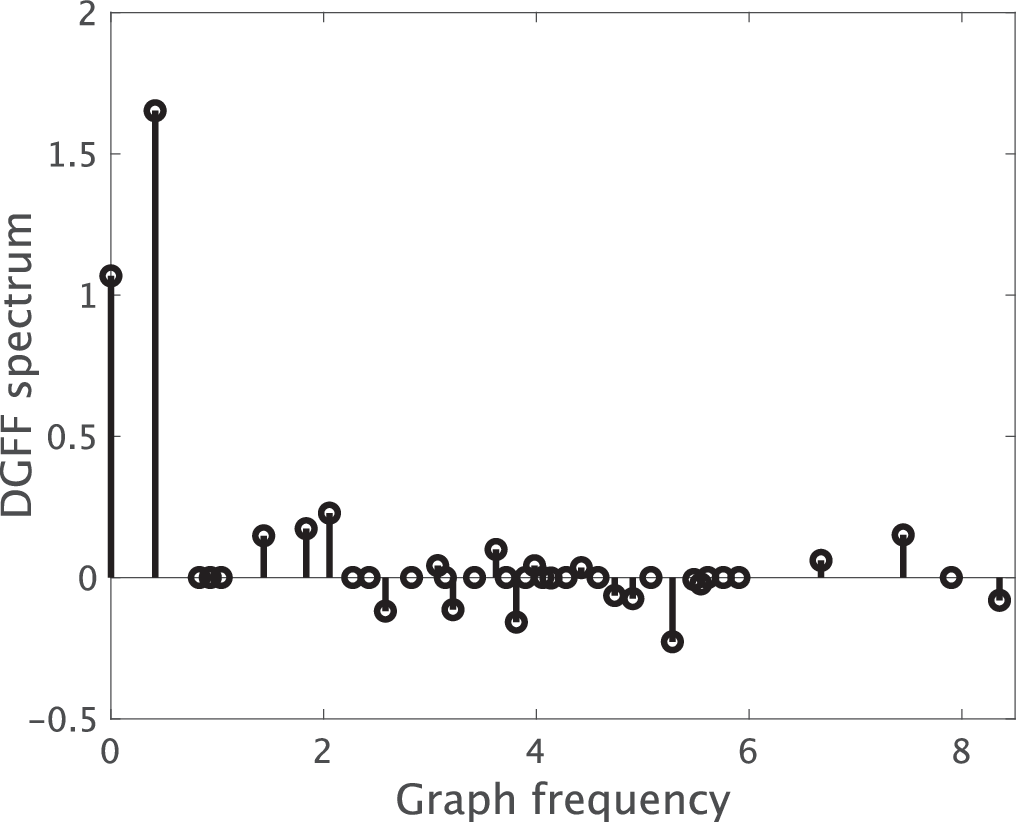}}
        \centerline{(c) Original signal spectrum}
    \end{minipage}
    \begin{minipage}[b]{0.49\linewidth}
    \centering
        \scalebox{0.23}{\includegraphics[keepaspectratio=true]{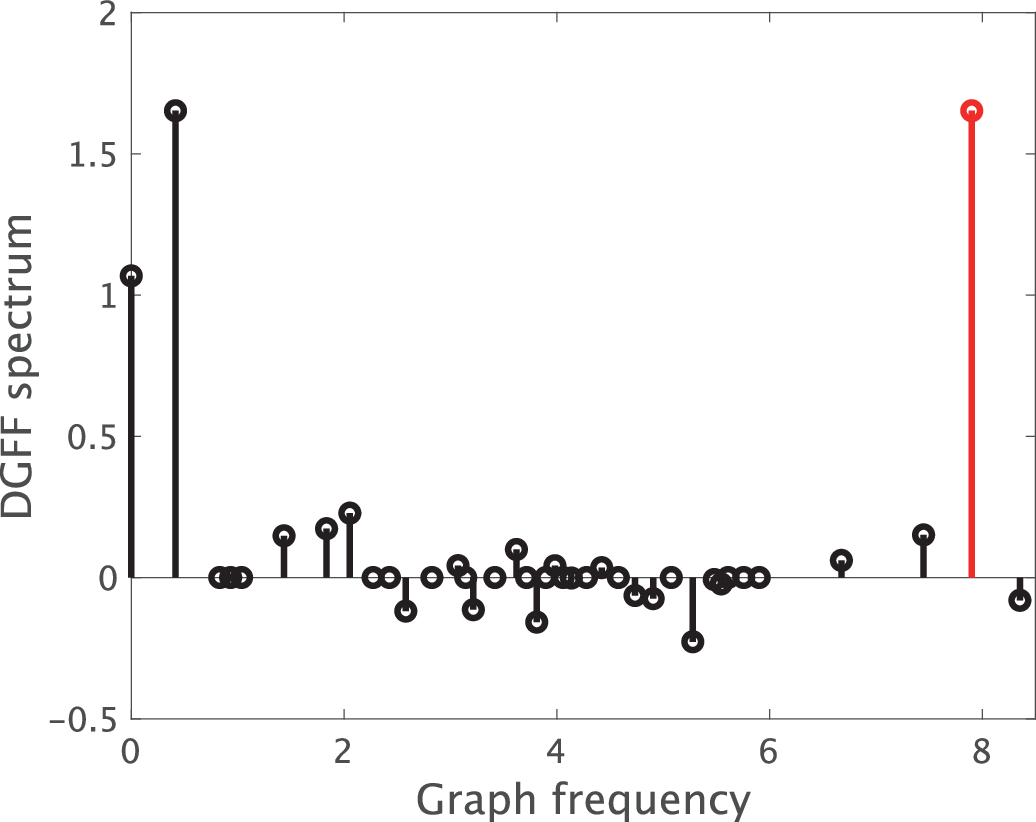}}
        \centerline{(d) Noise signal spectrum}
    \end{minipage}
\caption{(a) Random undirected graph signal (b) Noise signal generated by manipulating the original LiDGFF spectrum (c) LiDGFF spectrum of the original graph signal (d) LiDGFF spectrum of the generated noise signal (the red-colored coefficient is newly introduced into the original LiDGFF spectrum).}
\label{fig:demoRandamGraphSignal}
\vspace{-0.5cm}
\end{figure}

To demonstrate the effectiveness of DGS filtering, a simple experiment is conducted. A random graph signal $\mathbf{s}^{\star} \in \mathbb{R}^N$ is generated using a stochastic block model \cite{HOLLAND1983109} with two clusters, an internal edge probability of 0.7, and an external edge probability of 0.25. The signal values are set to 0.9 for one cluster and 0.1 for the other. An example is shown in Fig. \ref{fig:demoRandamGraphSignal}(a). Then, sparse coefficients are obtained using LiDGFF, $\mathbf{a}^{\star} = \mathcal{F}(\mathbf{s}^{\star}) \in \mathbb{R}^K$, as shown in Fig. \ref{fig:demoRandamGraphSignal}(c). Next, the maximum absolute value of $\mathbf{a}$, i.e., $\max({a_n^{\star}})$, is assigned to the highest intermediate graph frequency $\lambda_{n_\mathrm{z}}$, whose coefficient is zero, i.e., $n_\mathrm{z} = \max(\{n \ | \ a_{n}^{\star} = 0\})$ (the red-colored coefficient in Fig. \ref{fig:demoRandamGraphSignal}(d)). By multiplying the coefficients with the frame, the noisy signal $\mathbf{y}$ is generated as $\mathbf{y} = \mathbf{F}(\mathbf{a}^{\star} + \mathbf{n})$, where $\mathbf{n} \in \mathbb{R}^{K}$ is set as $\mathbf{n}_n = 0$ and $\mathbf{n}_{n_\mathrm{z}} = \max(\{a_n^{\star}\})$, shown in Fig. \ref{fig:demoRandamGraphSignal}(b).
This signal $\mathbf{y}$ is transformed into the frequency domain using GFB and LiDGFF, and filtering in \eqref{eq:DGFFfilter} is performed with a filter characterized by the frequency response $\widetilde{\mathbf{h}}_w := [ \widetilde{h}_1, \dots , \widetilde{h}_M ]^\top (\widetilde{h}_k = \mathbb{I}_{\{ k \le w \}})$, where $w$ corresponds to the cutoff frequency and $\mathbb{I}_{\{ k \le w \}} = 1$ for $k \leq w $, otherwise $\mathbb{I}_{\{ k \le w \}} = 0$, then finally the filtered signal $\widetilde{\mathbf{s}}$ is obtained. The relative recovery error $e_f/e$ \cite{DGFT} is used to evaluate the filter’s performance, where $e_f$ and $e$ are defined as $e_f = \| \widetilde{\mathbf{s}} - \mathbf{s}^{\star} \|_2 / \| \mathbf{s}^{\star} \|_2$ and $e = \| \mathbf{n} \|_2 / \| \mathbf{s}^{\star} \|_2$, respectively. 

\begin{figure*}[t]
\centering
    \begin{minipage}[b]{0.32\linewidth}
    \centering
        \scalebox{0.27}{\includegraphics[keepaspectratio=true]{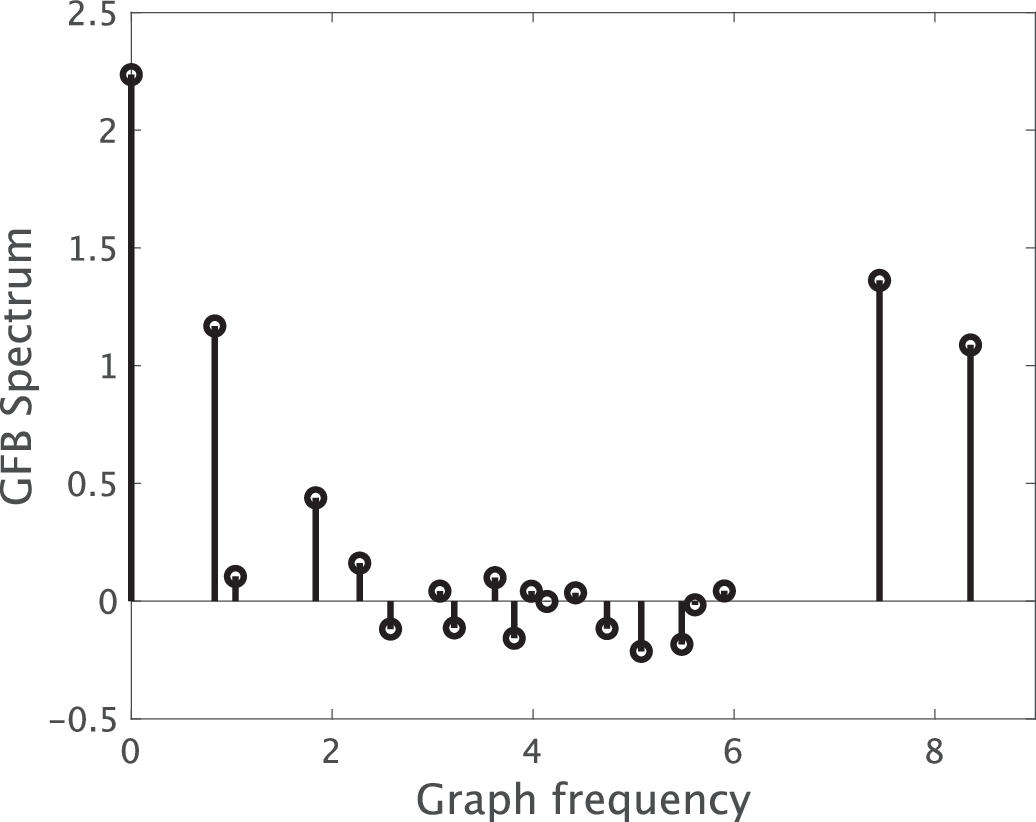}}
        \centerline{(a) GFB spectrum}
    \end{minipage}
    \begin{minipage}[b]{0.32\linewidth}
    \centering
        \scalebox{0.27}{\includegraphics[keepaspectratio=true]{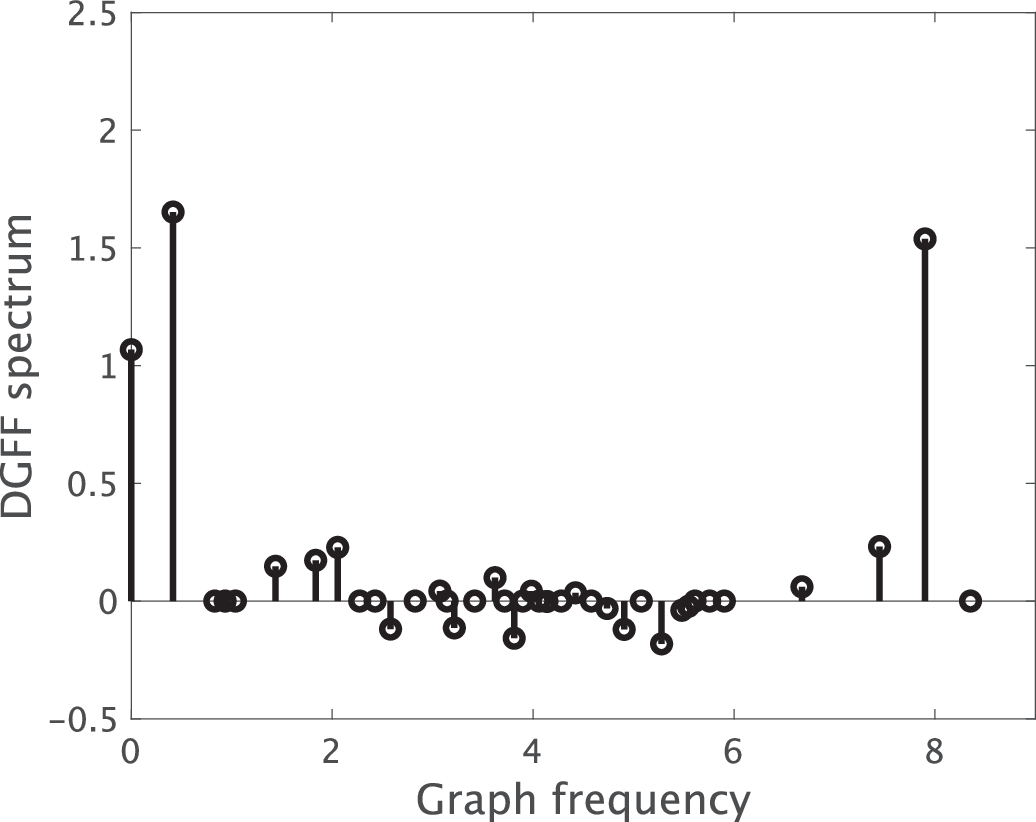}}
        \centerline{(b) LiDGFF spectrum}
    \end{minipage}
    \begin{minipage}[b]{0.32\linewidth}
    \centering
        \scalebox{0.27}{\includegraphics[keepaspectratio=true]{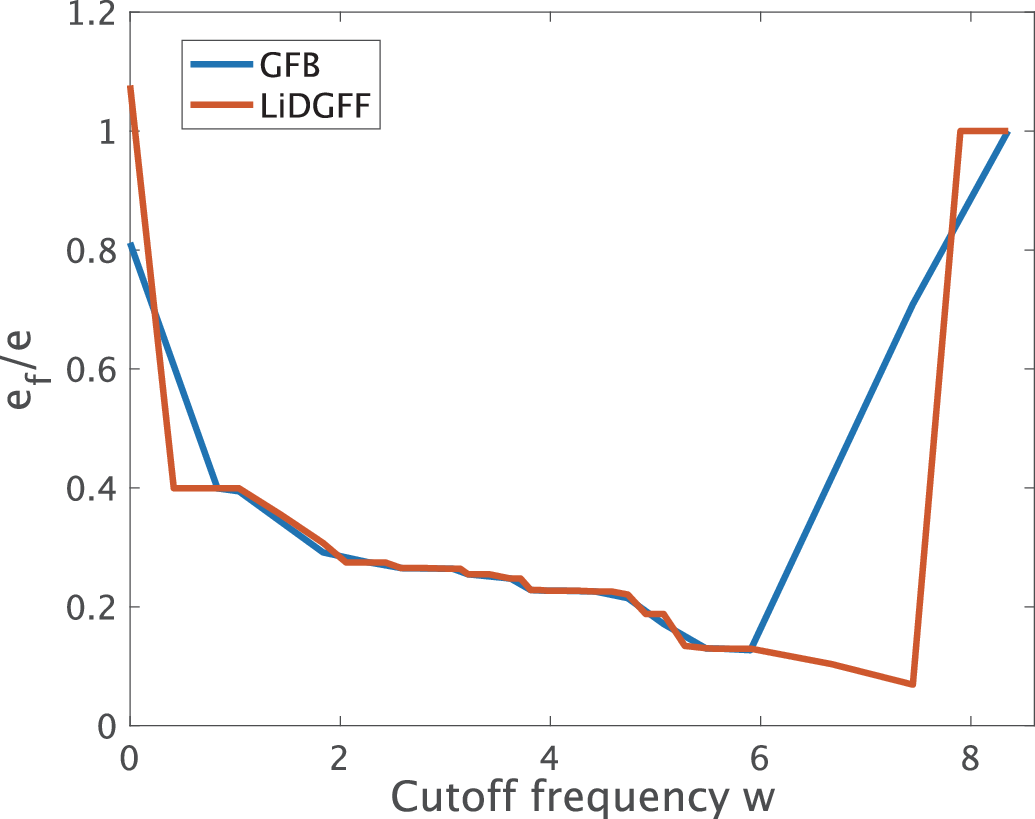}}
        \centerline{(c) Reconstruction error}
    \end{minipage}
\caption{(a), (b) Spectra obtained from the noisy signal shown in Fig. \ref{fig:demoRandamGraphSignal}(b) using GFB and LiDGFF (c) Reconstruction error $e_f / e$.}
\label{fig:demoFilteringResult}
\end{figure*}

The obtained spectra and reconstruction results are presented in Fig. \ref{fig:demoFilteringResult}. In the GFB spectrum, two elements corresponding to high-frequency components exhibit large values (Fig. \ref{fig:demoFilteringResult}(a)), whereas, in the LiDGFF spectrum, a single element holds a prominent value. The LiDGFF spectrum allows the noise component at intermediate graph frequencies to be represented with fewer frequency components, enabling the separation of the true signal components and noise components. This results in a smaller reconstruction error, as shown in Fig. \ref{fig:demoFilteringResult}(c). In contrast, GFB represents the noise components at intermediate graph frequencies as a linear combination of multiple frequency components. To remove the noise components, more frequency components must be suppressed, which also inadvertently suppresses parts of the true signal components. The proposed DGFF method can effectively represent components at intermediate graph frequencies, enabling precise filtering operations.

\subsubsection{Graph Signal Denoising by DGS Filtering}
\begin{figure}[t] 
  \centering 
 	  \scalebox{0.5}{\includegraphics[keepaspectratio=true]{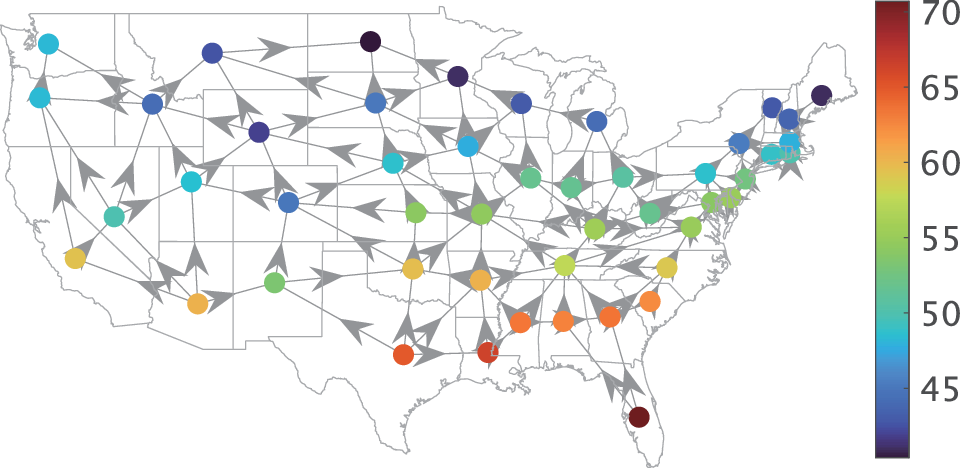}}
    \caption{Temperature signals obtained from the 48 contiguous states of the United States, excluding Alaska and Hawaii. Directed edges are determined based on latitude.} \label{fig:UStemp}
 \end{figure}

\begin{figure*}[t]
\centering
    \begin{minipage}[b]{0.32\linewidth}
    \centering
        \scalebox{0.27}{\includegraphics[keepaspectratio=true]{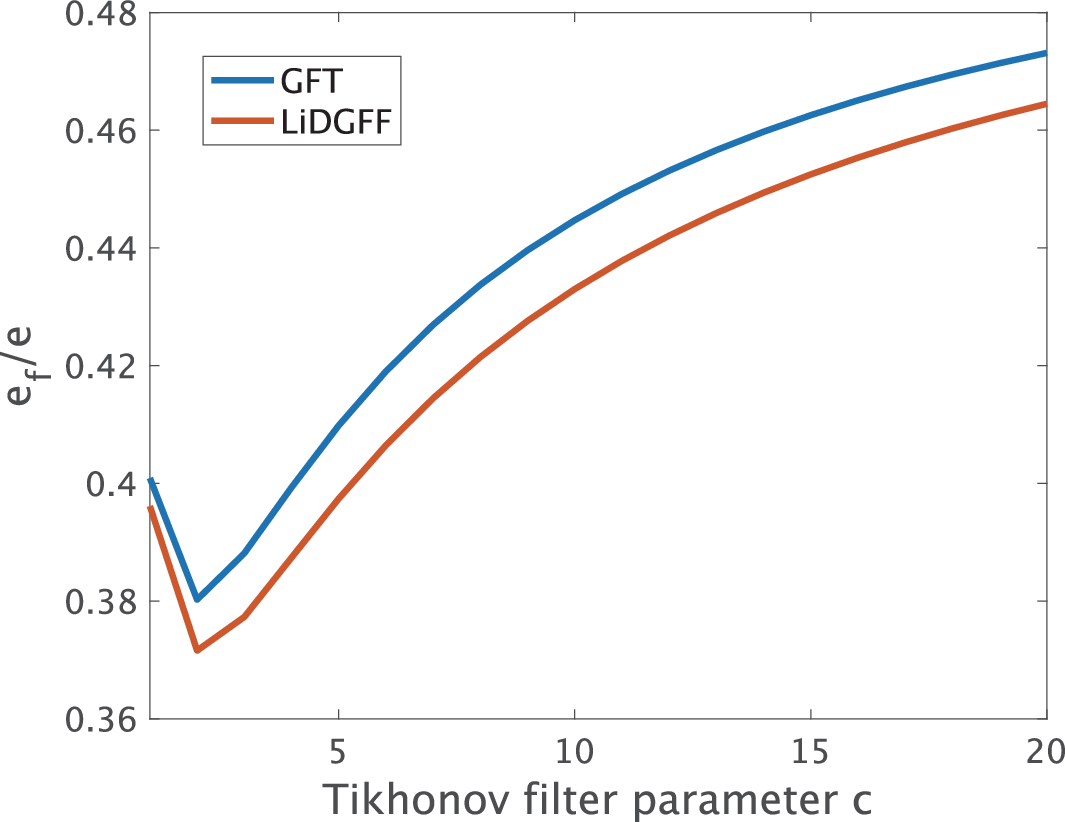}}
        \centerline{(a) GFB and LiDGFF}
    \end{minipage}
    \begin{minipage}[b]{0.32\linewidth}
    \centering
        \scalebox{0.27}{\includegraphics[keepaspectratio=true]{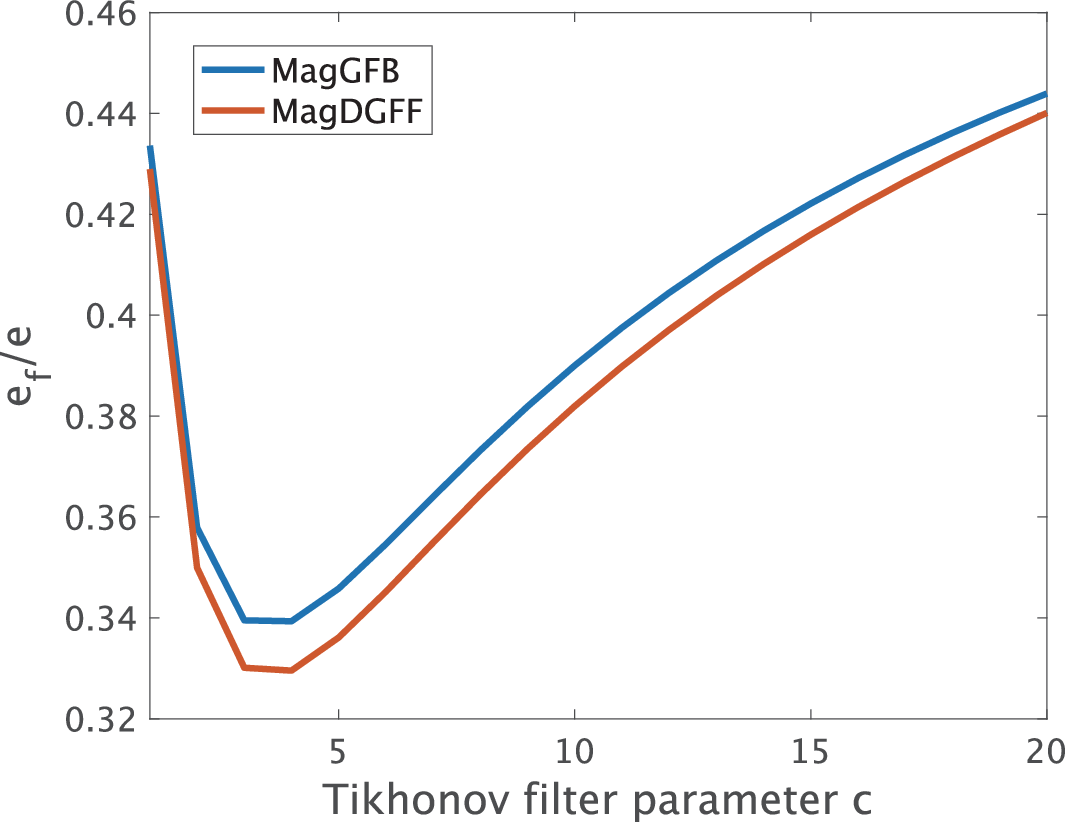}}
        \centerline{(b) MagGFB and MagDGFF}
    \end{minipage}
    \begin{minipage}[b]{0.32\linewidth}
    \centering
        \scalebox{0.27}{\includegraphics[keepaspectratio=true]{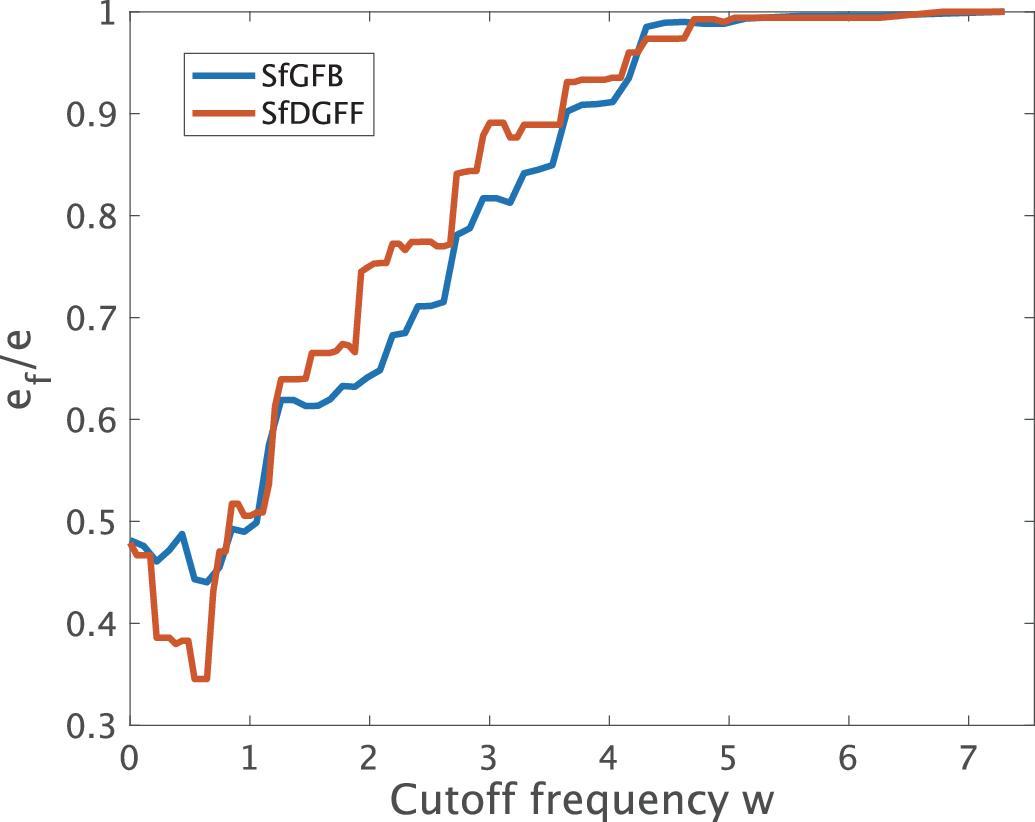}}
        \centerline{(c) SfGFB and SfDGFF}
    \end{minipage}
\caption{Reconstruction errors $e_f/e$ between the restored signals and the ground-truth signals after filtering. }
\label{fig:FilteringResults}
\end{figure*}

To evaluate DGS filtering, we perform noise removal and reconstruction on temperature signals over normalized undirected and directed USA graphs (see Fig. \ref{fig:UStemp}) \cite{MagGFT}. In this evaluation, the observed signal $\mathbf{y} = \mathbf{s}^{\star} + \mathbf{n} \in \mathbb{R}^N$ is used, where Gaussian noise $\mathbf{n}$ with a standard deviation of $\sigma = 0.2$ is added to the true signal $\mathbf{s}^{\star}$. For filtering undirected and directed graph signals based on the magnetic Laplacian, we employ a graph frequency domain filter using GFB/MagGFB and LiDGFF/MagDGFF filtering, as defined in \eqref{eq:DGFFcoef}. As a low-pass filter in \eqref{eq:DGFFfilter}, we adopt the Tikhonov filter $\widehat{\mathbf{h}}_{\boldsymbol{\Lambda}}=\begin{bmatrix}
    \frac{1}{1+c\lambda_1} & \cdots & \frac{1}{1+c\lambda_N}
\end{bmatrix}^\top $ \cite{MagGFT, dabush2024verifying, isufi2024graph, kalofolias2016learn}, where the parameter $c$ controls the attenuation of high-frequency components. The Tikhonov filter is originally related to signal denoising with GTV regularization, formulated as:
\begin{align} 
    \widetilde{\mathbf{y}} =&\ \argmin_{\mathbf{s} \in \mathbb{C}^{N} } \frac{1}{2}\|\mathbf{s}- \mathbf{y}\|_2^2 + c \mathbf{s}^\top\mathbf{L}\mathbf{s}\nonumber \\
    = &\  (\mathbf{I} + c\mathbf{L})^{-1} \mathbf{y} = \mathbf{U}\mathrm{Diag}\left(\widehat{\mathbf{h}}_{\boldsymbol{\Lambda}}\right)\mathbf{U}^\top\mathbf{y},
\end{align}
where $\mathbf{L} \in \mathbb{R}^{N\times N}$ is a (Hermitian) symmetric graph Laplacian, and $\mathbf{U}$ represents its GFB. For DGFFs, the filtered signal is obtained as $\widetilde{\mathbf{y}} = \mathbf{F}\mathrm{Diag}\left(\widehat{\mathbf{h}}_{\boldsymbol{\Lambda}}\right)\mathcal{F}(\mathbf{y})$, where $\widehat{\mathbf{h}}_{\boldsymbol{\Lambda}} = \begin{bmatrix}
    \frac{1}{1+c\lambda_1} & \cdots & \frac{1}{1+c\lambda_M} 
\end{bmatrix}^\top$. For directed graph signals, a comparison between the SfGFB and SfDGFF is also performed. Note that the Tikhonov filter is not suitable for GS and DGS filtering based on SfGFB and SfDGFF, as they are derived from DV rather than GTV. Thus, according to \cite{DGFT}, we simply applied the same graph spectral filter $\widetilde{\mathbf{h}} = [ \widetilde{h}_1, \dots , \widetilde{h}_M ]^\top (\widetilde{h}_k = \mathbb{I}_{\{ k \le w \}})$ which was also used in Sec.\ref{subsec:SGFDGS}. To assess the filter's performance, we use the reconstruction error $e_f/e$. Note that for the MagGFB and MagDGFF, where the output is complex-valued, the error was calculated using only the real part. 
The results of signal reconstruction through filtering are shown in Fig. \ref{fig:FilteringResults}. For filtering based on both undirected graphs and the magnetic Laplacian, the LiDGFF and MagDGFF filtering consistently outperformed the GFB and MagGFB filtering for each cutoff frequency. In the case of DGFF filtering, the selection of sparse coefficients allowed the filter to focus on the most significant frequency components of the signal, resulting in more precise filtering and more effective noise reduction than GFB. Additionally, experiments using SfGFB and SfDGFF with different filter designs demonstrated that the SfDGFF filtering based on sparse coefficients also exhibited more effective noise reduction performance than the SfGFB filtering.

\subsection{Graph Signal Recovery by Sparse Representation}

\begin{table}[t]
  \centering
  \caption{Number of Vectors For Each Basis and Frame}
  \renewcommand{\arraystretch}{1.2}
  \setlength{\tabcolsep}{4pt} 
  \scalebox{0.97}{
    \centering
\begin{tabular}{c|c|c|c|c} 
\thline
Undirected Graph   &  U.S. temp & Community & Swiss Roll & Minnesota \\ \hline
Nodes & 48 & 250  & 1000  & 2642 \\ \hline \hline
GFB    & 48 & 250 & 1000 & 2642  \\
RGFF   &  96 & 500 & 2000 & 5284 \\ 
LiDGFF     &  95 & 499 & 1999 & 5283  \\ 
lrLiDGFF   &   88 & 455 & 1632  & 4649  \\ \thline
\end{tabular}}
\\
  \scalebox{0.97}{
    \centering
  \begin{tabular}{c|c|c} 
\multicolumn{3}{c}{} \\ \thline
Directed  Graph  &  France temp  & U.S. temp \\ \hline
Nodes &  32 & 48  \\ \hline \hline
MagGFB    &  32 & 48 \\
MagDGFF   & 63 & 95  \\ 
SfGFB     & 32 & 48  \\ 
SfDGFF   & 63 & 95 \\ \thline
\end{tabular}
}
\label{tab:NumOfNodes}
\end{table}
In this section, we evaluate the proposed frames in graph signal recovery by SR. The undirected graph signals used in the experiments include temperature signals on a symmetrized graph of United States \cite{MagGFT}, community graph signals \cite{Unrolling}, Swiss Roll graph signals \cite{TFfilter}, and the Minnesota traffic graph \cite{OSGFB}. For directed graph signals, we use U.S. temp \cite{MagGFT} and temperature signals from 32 monitoring stations in France \cite{FranceTemp, cheng2022}. The direction of edges in the graph for the French temperature signals is assigned from nodes with lower latitude to nodes with higher latitude, following the methodology in \cite{MagGFT, DGFT, Mohammad2024}.
We compared the proposed LiDGFF and lrLiDGFF with the undirected GFB and the RGFT-based frame (hereafter RGFF) \cite{RGFT}. For directed graph signals, we use the MagGFB, MagDGFF, SfGFB, and SfDGFF. The number of nodes for each graph and the number of vectors for each basis and frame are summarized in Table \ref{tab:NumOfNodes}. 
The parameter $\rho$ for RGFF (see Sec. \ref{subsubsec:GFFDict}) was set to $\rho = \frac{1}{4} \min_{k=1}^{N-1} \{ \lambda_{k+1} - \lambda_{k} \}$. Additionally, the encoding parameters for MagGFB and MagDGFF (i.e., $q$ in \eqref{eq:maggfb}) are set to 0.01. 

Table \ref{tab:SfIMerror} presents the error between the ideal intermediate graph frequencies and those derived from the SfDGFF, computed as: 
\begin{align} \sqrt{\sum_{k=1}^{N-1} \left(\mathcal{DV}(\mathbf{u}_k^{(\alpha, \beta)}) - \left(\alpha\mathcal{DV}(\mathbf{u}_k) + \beta\mathcal{DV}(\mathbf{u}_{k+1})\right)\right)^2} ,\end{align} for $(\alpha, \beta) = (0.5,0.5)$. The results demonstrate that, under the parameter settings employed in this evaluation, the derived intermediate graph frequencies closely align with the ideal intermediate graph frequencies with high accuracy.
\begin{table}[t]
  \centering
  \caption{Comparison of Errors in SfDGFF Intermediate Graph Frequencies with Ideal Values}
  \renewcommand{\arraystretch}{1.2}
  \setlength{\tabcolsep}{4pt} 
  \scalebox{0.97}{
    \centering
\begin{tabular}{c|c} 
\thline
Graph   &  Error \\ \hline
Synthetic directed graph (Fig. \ref{fig:UDSG}) & $4.55\times10^{-11}$ \\
U.S. temp    & $1.98\times10^{-08}$ \\
France temp   &  $1.81\times10^{-07}$ \\ 
\thline
\end{tabular}}
\label{tab:SfIMerror}
\end{table}

We first consider the problem of reconstructing a true signal $\mathbf{s}^\star \in \mathbb{R}^N$ from a randomly missing signal in the absence of noise. The observed signal, $\mathbf{y} \in \mathbb{R}^{pN}$, is obtained using the sampling matrix $\mathbf{\Phi}\in \{0,1\}^{pN \times N}$ as $\mathbf{y} = \mathbf{\Phi} \mathbf{s}^{\star}$, where the parameter $p \in (0,1)$ corresponds to the sampling rate. In the graph signal recovery problem, we estimate the true signal $\widetilde{\mathbf{x}}$ by finding $\widetilde{\mathbf{a}}$ based on the following equation:
\begin{align}\label{eq:Inpainting}
       \widetilde{\mathbf{a}} = \argmin_{\mathbf{a} \in \mathbb{C}^K} \|\mathbf{a}\|_1\ \mathrm{s.t.}\  \mathbf{\Phi}\mathbf{F}\mathbf{a} = \mathbf{y} \ \Longrightarrow \widetilde{\mathbf{s}} = \mathbf{F}\widetilde{\mathbf{a}}.
\end{align}
Here, the true signal $\mathbf{s}^{\star}$ is assumed to lie within the range $[-1, 1]^N$. Evaluations were conducted at the sampling rates of $30\%$ and $70\%$ ($p= 0.3, 0.7$). In addition, we consider graph signal recovery in the presence of additive Gaussian noise using SR as in the following optimization problem:
\begin{align} \label{eq:Interpolation}
\widetilde{\mathbf{a}} = \argmin_{\mathbf{a} \in \mathbb{C}^K} \|\mathbf{a}\|_1\ \mathrm{s.t.}\ \|\mathbf{\Phi}\mathbf{F}\mathbf{a} - \mathbf{y}\|_2 \leq \epsilon\ \Longrightarrow \widetilde{\mathbf{s}} = \mathbf{F}\widetilde{\mathbf{a}}.
\end{align}
In this problem, the true signal is estimated from the observed signal in the presence of Gaussian noise, $\mathbf{y} = \mathbf{\Phi} \mathbf{s}^{\star} + \mathbf{n}$, where the sampling rate is $p \in \{0.3, 0.7 \}$, and the standard deviations of Gaussian noise are set as $\sigma \in \{0.05, 0.1\}$. Additionally, the $\ell_2$-norm constraint parameter $\epsilon$ is set as $0.90\sigma\sqrt{N}$. In this study, we employed primal-dual splitting algorithms \cite{PDS1, PDS2, PDS3} to solve equations (\ref{eq:Inpainting}) and (\ref{eq:Interpolation}) (for more details, see Appendix \ref{sec:AIR}).

The reconstruction results (SNR [dB]) for undirected graph signals are shown in Tables \ref{tab:UDGSInpainting2} and \ref{tab:UDGSInterpolation}. The performance of signal reconstruction using SRs on undirected graphs is discussed as follows. First, the reconstruction errors of the RGFF are almost the same as those of the GFB since the RGFF exhibited a graph frequency distribution nearly identical to that of GFB. The parameter $\rho$ for the RGFF, which is determined based on the minimum interval of the original eigenvalues, is likely close to zero $\rho \approx 0$, especially in large scale graph signals. This results in the modified graph Laplacian are nearly identical to $\mathbf{L} - 2\rho \mathbf{I} \approx \mathbf{L}$, and thus the additional eigenvectors $\{\mathbf{v}_k\}_{k=1}^{N}$ for the RGFF are the copies of the original eigenvectors $\{\mathbf{u}_k\}_{k=1}^{N} \approx \{\mathbf{v}_k\}_{k=1}^{N}$. In contrast, the proposed LiDGFF and lrLiDGFF successfully generate intermediate graph frequency vectors for each graph and consistently demonstrate superior reconstruction performance in most cases.

\begin{table}[!tb]
  \centering
  \caption{Undirected Graph Signal Recovery in the Absence of Noise by SR (SNR [dB])}
  \renewcommand{\arraystretch}{1.2}
  \setlength{\tabcolsep}{4pt} 
  \scalebox{0.95}{
\begin{tabular}{c|cc|cc|cc|cc} 
\thline
   & \multicolumn{2}{c|}{U.S. temp}  & \multicolumn{2}{c|}{Community} & \multicolumn{2}{c|}{Swiss Roll} & \multicolumn{2}{c}{Minnesota} \\ \hline
Sampling Rate & 70\% & 30\% & 70\%  & 30\% & 70\%  & 30\% & 70\%  & 30\% \\ \hline \hline
Observation    & 5.35 & 1.49 & 5.08 & 1.54 & 5.32 & 1.56 & 5.22 & 1.54  \\ \hline
GFB    & 25.06 & 16.56 & 22.01 & 17.76 & 27.53 & 21.05 & 18.15 & 11.94 \\
RGFF   & 25.06 & 16.56 & 22.01 & 17.76 & 27.53 & 21.05 & 18.14 & 11.93 \\ 
LiDGFF     & \textbf{26.06} & \textbf{17.35} & 22.37 & \textbf{18.12} & \textbf{27.99} & \textbf{21.75} & \textbf{18.44} & \textbf{12.35} \\ 
lrLiDGFF   & 25.94 & 17.32 & \textbf{22.55} & 18.10 & 27.45 & 21.21 & 18.35 & 12.14 \\ \thline
\end{tabular}}
\label{tab:UDGSInpainting2}
 \vspace{0.25cm} 
  \centering
  \caption{Undirected Graph Signal Recovery in the Presence of Noise by SR (SNR [dB])}
  \renewcommand{\arraystretch}{1.2}
  \setlength{\tabcolsep}{4pt} 
  \scalebox{0.95}{
\begin{tabular}{c|cc|cc|cc|cc} 
\thline
   & \multicolumn{2}{c|}{U.S. temp}  & \multicolumn{2}{c|}{Community} & \multicolumn{2}{c|}{Swiss Roll} & \multicolumn{2}{c}{Minnesota} \\ \hline
   Sampling Rate & \multicolumn{8}{c}{70\%} \\ \hline 
Noise ($\sigma$) & $0.05$ & $0.1$ & $0.05$ & $0.1$ & $0.05$ & $0.1$ & $0.05$ & $0.1$ \\ \hline \hline
Observation    & 5.28  & 5.03  & 5.28  & 5.09  & 5.20  & 5.07  & 5.20  & 5.12  \\ \hline
GFB            & 21.48 & 18.66 & 19.75 & 17.96 & 24.10 & \textbf{21.07} & 17.57 & 16.16  \\ 
RGFF           & 21.48 & 18.66 & 19.75 & 17.96 & 24.10 & \textbf{21.07} & 17.57 & 16.16 \\ 
LiDGFF   & 21.65 & 18.90 & \textbf{19.98} & \textbf{17.97} & \textbf{24.14} & 20.98 & 17.58 & 15.88 \\ 
lrLiDGFF  & \textbf{21.81} & \textbf{18.96} & 19.92 & \textbf{17.97} & 23.95 & 20.84 & \textbf{17.79} & \textbf{16.31} \\ \hline\hline
   Sampling Rate & \multicolumn{8}{c}{30\%} \\ \hline 
Noise ($\sigma$) & $0.05$ & $0.1$ & $0.05$ & $0.1$ & $0.05$ & $0.1$ & $0.05$ & $0.1$ \\ \hline \hline
Observation    & 1.38  & 1.52  & 1.43  & 1.49  & 1.55  & 1.49  & 1.54  & 1.53  \\ \hline
GFB            & 15.30 & 15.96 & 16.92 & 15.59 & 20.23 & 18.44 & 11.77 & 11.50  \\ 
RGFF           & 15.30 & 15.96 & 16.92 & 15.59 & 20.23 & 18.44 & 11.77 & 11.50 \\ 
LiDGFF   & 16.34 & 16.24 & 17.21 & 15.70 & \textbf{20.59} & \textbf{18.67} & \textbf{12.17} & \textbf{11.96} \\ 
lrLiDGFF  & \textbf{16.39} & \textbf{16.28} & \textbf{17.22} & \textbf{15.73} & 20.18 & 18.40 & 11.94 & 11.74 \\ \thline
\end{tabular}
}
\vspace{-0.15cm}
\label{tab:UDGSInterpolation}
\end{table}
 
The reconstruction results for directed graph signals are shown in Table \ref{tab:DGSInpainting}. For the MagGFB and MagDGFF, where the output is complex-valued, the SNR was calculated using only the real part. In directed graph signals as well, the DGFF methods extended from each GFB approach demonstrated superior reconstruction performance due to their denser graph frequencies and SR efficiency.
\begin{table}[!tb]
  \centering
  \caption{Directed Graph Signal Recovery by SR : (SNR [dB])}
  \renewcommand{\arraystretch}{1.2}
  \setlength{\tabcolsep}{4pt} 
    \scalebox{0.95}{
\begin{tabular}{c|ccc|ccc} 
\thline
   & \multicolumn{6}{c}{France temp} \\ \hline
Sampling Rate &  \multicolumn{3}{c|}{70\% }  &  \multicolumn{3}{c}{30\% }   \\ \hline 
Noise ($\sigma$) & --- & 0.05 &  0.1  & --- & $0.05$ & $0.1$ \\ \hline \hline
Observation   & 4.96 & 5.09 & 4.58 & 1.51 & 1.61 & 1.57  \\ \hline
MagGFB & 22.51 & 21.18 & 17.87 & 12.99 & \textbf{13.60} & \textbf{13.32} \\ 
MagDGFF  & \textbf{24.41} & \textbf{22.31} & \textbf{18.81} & \textbf{14.53} & 13.11 & 12.81 \\ \hline
SfGFB & 18.69 & 18.68 & 15.86 & 12.84 & \textbf{13.28} &  \textbf{12.82} \\ 
SfDGFF  & \textbf{20.00} & \textbf{18.84} & \textbf{16.01} & \textbf{13.06} & 12.98 & 12.44 
\\ \thline
 \multicolumn{7}{c}{} \\\thline
   &  \multicolumn{6}{c}{U.S. temp} \\ \hline
Sampling Rate &  \multicolumn{3}{c|}{70\% }  &  \multicolumn{3}{c}{30\% }   \\ \hline 
Noise ($\sigma$)  & --- & 0.05 &  0.1  & --- & $0.05$ & $0.1$  \\ \hline \hline
Observation   & 5.48 & 5.25 & 5.16 & 1.53 & 1.50 & 1.49 \\ \hline
MagGFB & 26.96 & 22.51 & 19.22 & 17.04 & 16.63 & 15.03 \\ 
MagDGFF  & \textbf{27.32} & \textbf{22.89} & \textbf{19.39} & \textbf{17.73} & \textbf{17.04} & \textbf{15.58} \\ \hline
SfGFB & 21.66 & 20.16 & 17.72 & 16.83 & \textbf{15.91} & 15.48 \\ 
SfDGFF  & \textbf{24.18} & \textbf{20.73} & \textbf{17.81} & \textbf{17.15} & 15.83 & \textbf{15.64} 
\\ \thline
\end{tabular}
}
\label{tab:DGSInpainting}
\end{table}

\subsubsection{Comparison of Computational Time and Iteration}
\begin{figure}[t]
\centering
    \begin{minipage}[b]{0.48\linewidth}
    \centering
        \scalebox{0.165}{\includegraphics[keepaspectratio=true]{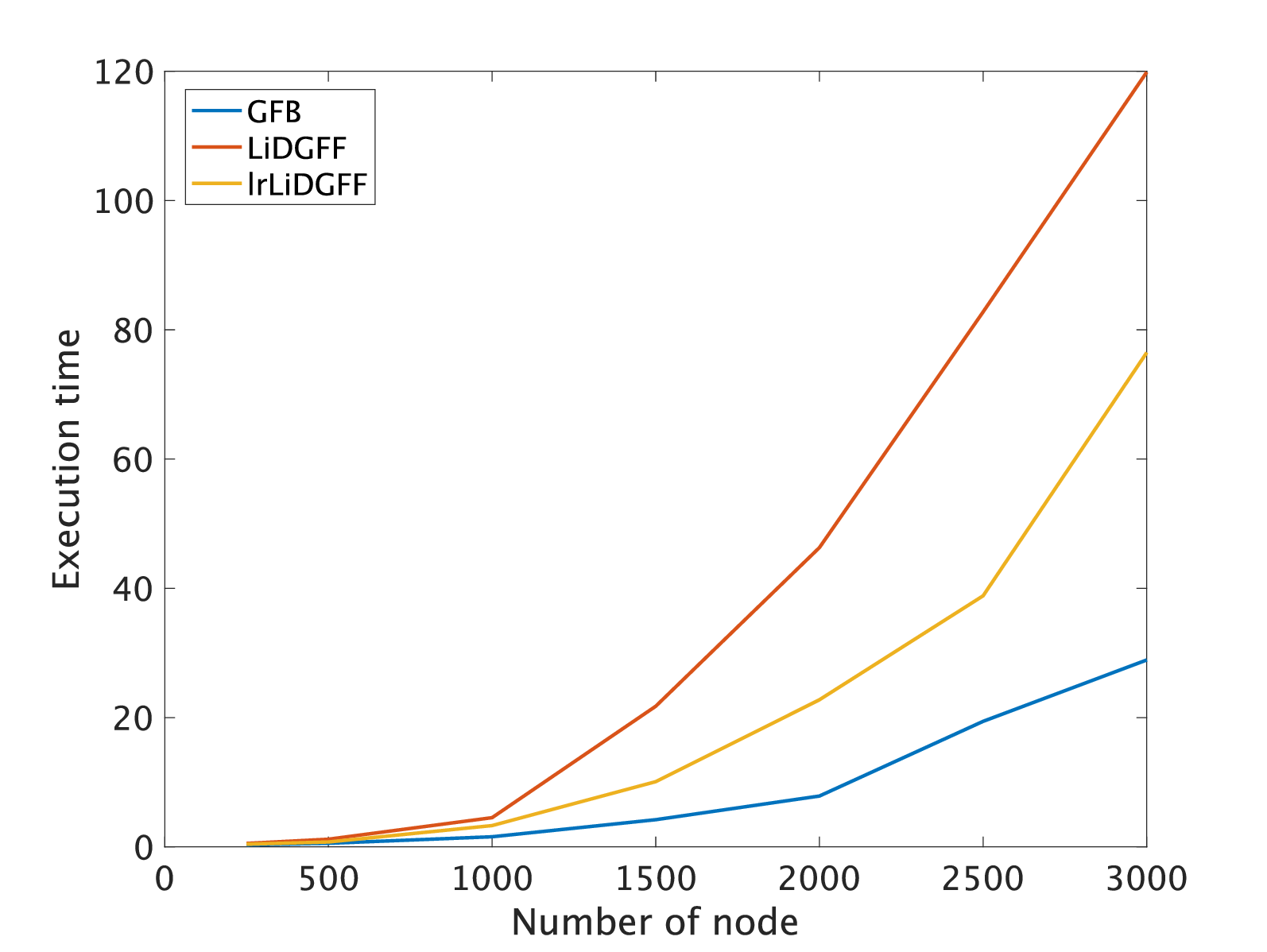}}
        \centerline{(a) Computational time}
    \end{minipage}
    \begin{minipage}[b]{0.48\linewidth}
    \centering
        \scalebox{0.165}{\includegraphics[keepaspectratio=true]{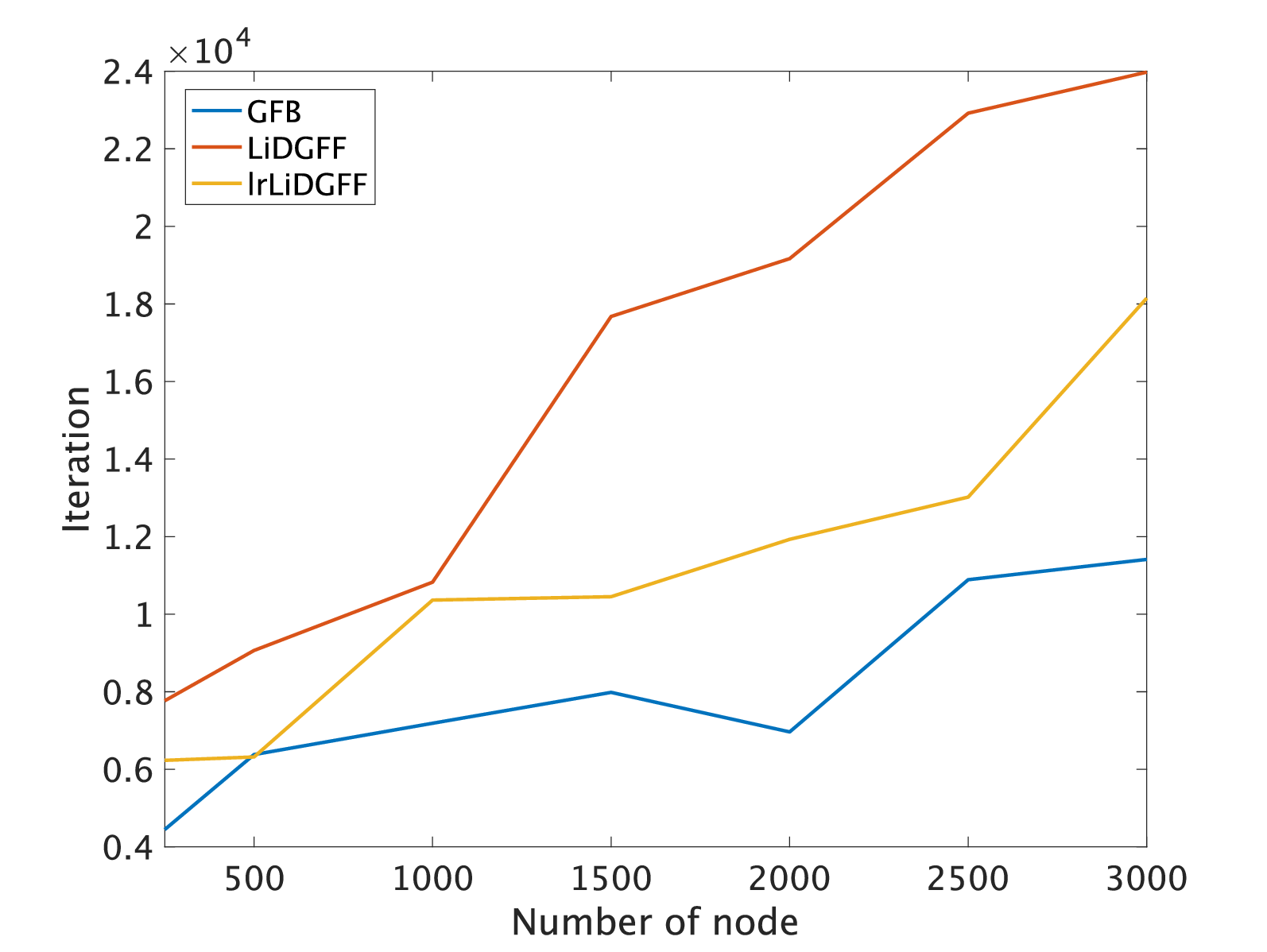}}
        \centerline{(b) Iteration}
    \end{minipage}
\caption{Computational time to convergence and the number of iterations for SR using GFB, LiDGFF, and lrLiDGFF as the size of the Swiss Roll graph increases to $\{250, 500, 1000, 1500, 2000, 2500, 3000\}$. The number of frequency components for each method is $N$ for GFB, $2N-1$ for LiDGFF, and $\{424, 811, 1632, 2426, 3244, 4036, 4893\}$ for lrLiDGFF.}
\vspace{-0.5cm}
\label{fig:ComputePlot}
\end{figure}

The computational costs and the number of iterations for the GFB, the LiDGFF, and the lrLiDGFF were also compared. In this evaluation, the computational time and iteration required to find the optimal solution of the SR problem \eqref{eq:Interpolation} was measured while gradually increasing the number of nodes in a Swiss Roll graph \cite{gspbox}. Here, each algorithm was executed until the stopping criterion (see Appendix \ref{sec:AIR}) was satisfied. The results are shown in Fig. \ref{fig:ComputePlot}.
As can be seen in the figure, the LiDGFF requires significantly more computational time than the GFB as the number of nodes increases, due to its retention of a large number of frequency components. In contrast, the lrLiDGFF employs a method that measures the frequency spacing of the GFB and selectively adds frequency components, resulting in fewer frequency components compared to the LiDGFF. Consequently, the lrLiDGFF demonstrates reduced computational time as the number of nodes increases. Similarly, due to the richness of frequency components, the LiDGFF requires more iterations than the GFB, whereas the lrLiDGFF achieves a reduction in the number of iterations compared to the LiDGFF.

\section{Concluding Remark}\label{sec:conclusion}
This paper proposed the DGFFs, low-redundant DGFFs, and DGS filtering based on the DGFFs to enhance the analysis and processing of undirected and directed graph signals. These frameworks enable more detailed and less biased analysis in the graph spectral domain for a wide range of graphs, from small-scale to large-scale. Furthermore, the superior performance of DGS filtering in signal denoising experiments suggests its potential to improve other tasks in graph spectral processing. Additionally, the introduction of redundancy enhances SR in the graph spectral domain, leading to improved signal reconstruction accuracy.

\appendices
\section{Detailed Algorithms for \eqref{eq:DGFFcoef}, \eqref{eq:Inpainting}, and \eqref{eq:Interpolation} }\label{sec:AIR}
To solve \eqref{eq:DGFFcoef}, \eqref{eq:Inpainting}, and \eqref{eq:Interpolation}, the primal-dual splitting algorithm \cite{PDS1, PDS2, PDS3} is used. Consider the following convex optimization problem to find
\begin{align}
\label{eq:pds}
\mathbf{x}^{\star} \in \argmin_{\mathbf{x}\in \mathbb{R}^{N_1}} f(\mathbf{x}) + g(\mathbf{A}\mathbf{x}),
\end{align}
where $f\in \Gamma_0(\mathbb{R}^{N_1})$, $g \in \Gamma_0(\mathbb{R}^{N_2})$ ($\Gamma_0(\mathbb{R}^{N_2})$ is the set of proper lower semicontinuous convex functions \cite{Bauschke2011} on $\mathbb{R}^{N}$), and $\mathbf{A} \in \mathbb{R}^{{N_2}\times {N_1}}$. Then, the optimal solution $\mathbf{x}^{\star}$, can be obtained as:
\begin{align}
\label{eq:pdsalg}
\begin{cases}
\mathbf{x}^{(i+1)}:= \mathrm{prox}_{\gamma_1 f} (\mathbf{x}^{(i)} - \gamma_1 \mathbf{A}^{\top}\mathbf{z}^{(i)}) \\
\mathbf{z}^{(i+1)}:= \mathrm{prox}_{\gamma_2 g^{\ast}} (\mathbf{z}^{(i)} + \gamma_2 \mathbf{A}(2\mathbf{x}^{(i+1)} - \mathbf{x}^{(i)}))
\end{cases},
\end{align}
where $\mathrm{prox}$ denotes the {proximal operator}\footnote{The proximity operator, $\mathrm{prox}_{\gamma f}:\mathbb{R}^{N} \rightarrow \mathbb{R}^{N}$, is defined for a function $f \in \Gamma_0(\mathbb{R}^N)$ and an index $\gamma \in (0, \infty)$ by \cite{Bauschke2011} 
$
	\mathrm{prox}_{\gamma f}(\mathbf{x}) := \argmin_{\mathbf{y}\in \mathbb{R}^N} \gamma f(\mathbf{y}) + \frac{1}{2} \|\mathbf{x}-\mathbf{y}\|^2_2.
$} \cite{Bauschke2011}, $g^{\ast}$ is the conjugate function\footnote{For $\forall f \in \Gamma_0(\mathbb{R}^p)$, the conjugate function $f^\ast$ of $f$ is defined as
$
f^\ast({\bm \xi}) := \sup_{\mathbf{x} \in \mathbb{R}^N} \langle \mathbf{x}, {\bm \xi } \rangle - f(\mathbf{x}),
$
and its proximity operator is calculated as:
$
\mathrm{prox}_{\gamma f^\ast}(\mathbf{x}) = \mathbf{x} - \gamma\mathrm{prox}_{\frac{1}{\gamma}f} \left(\frac{1}{\gamma}\mathbf{x}\right)
$.
} \cite{Bauschke2011} of $g$.  In the experiments, the parameters $\gamma_1$ and $\gamma_2$ in \eqref{eq:pdsalg}, are chosen as 0.01 and $\frac{1}{12\gamma_1}$. For each problem in \eqref{eq:DGFFcoef}, \eqref{eq:Inpainting}, and \eqref{eq:Interpolation}, the functions $f$ and $g$, and the matrix $\mathbf{A}$ are set as $f(\mathbf{x}) = \|\mathbf{x}\|_1$, $g(\mathbf{A}\mathbf{x}) = \iota_{\mathcal{B}_2(\mathbf{y},\epsilon)}(\mathbf{A}\mathbf{x})$, and
\begin{align} 
\label{eq:setting}
\begin{cases}
\epsilon = 0,\ \mathbf{A} = \mathbf{I} & (\mathrm{for}\ \eqref{eq:DGFFcoef})   \\
\epsilon = 0,\ \mathbf{A} = \boldsymbol{\Phi}\mathbf{F} & (\mathrm{for}\ \eqref{eq:Inpainting})   \\
\epsilon = 0.90\sigma\sqrt{N},\ \mathbf{A} = \boldsymbol{\Phi}\mathbf{F} & (\mathrm{for}\ \eqref{eq:Interpolation})   
\end{cases}.
\end{align}
Under the setting, the optimal solution can be obtained by iterating the steps in \eqref{eq:pdsalg}\footnote{For $\mathbf{x} \in \mathbb{R}^N$, $[\mathrm{prox}_{\gamma \|\cdot \|_{1}}(\mathbf{x})]_i  = \mathrm{sign}(x_i) \max \{|x_i|-\gamma,0\}$ (soft-thresholding), the projection onto the $\ell_2$-norm ball $\mathcal{B}_2(\mathbf{y},\epsilon)$ and $\mathfrak{C}_2 =  \{ \mathbf{x} \ |\ \|\mathbf{x} - \mathbf{y} \|_2 \leq \epsilon  \}$ is $\mathrm{prox}_{\iota_{\mathcal{B}_2(\mathbf{y},\epsilon)}}[\mathbf{x}] = \mathbf{y} + \epsilon \frac{\mathbf{x}-\mathbf{y}}{\|\mathbf{x}-\mathbf{y}\|_2} $, where $\mathbf{y} \in \mathbb{R}^N$ is an observation. }. The stopping criterion is $\|\mathbf{x}^{(i+1)}-\mathbf{x}^{(i)}\|_2\leq 10^{-12}$. 

\bibliographystyle{IEEEtran}
\bibliography{refs}

\end{document}